\documentclass[aps,prb,reprint,showpacs,longbibliography]{revtex4-2}
\usepackage{graphicx} 
\usepackage{amsmath} 
\usepackage{mathtools} 
\usepackage{gensymb} 
\usepackage{textcomp}
\usepackage{bm} 
\usepackage{natbib}
\usepackage{natmove}
\usepackage[colorlinks=true, linkcolor=blue, citecolor=blue, urlcolor=blue, linktoc=page, bookmarks=false, pdfstartview={FitH}, pdfborder={0 0 0.0 [3 3]}]{hyperref} 
\usepackage{amssymb}
\usepackage{color} 
\usepackage{dcolumn} 
\newcolumntype{.}{D{.}{.}{2.1}}
\newcolumntype{-}{D{.}{.}{4.0}}
\usepackage{makecell}
\usepackage{multirow}
\usepackage{cleveref}
\usepackage{easyReview} 
\crefname{figure}{Fig.}{Figs}
\Crefname{figure}{Figure}{Figures}
\crefname{table}{Table}{Tables}
\crefname{equation}{Eq.}{Eqs.}
\crefname{section}{Sec.}{Secs.}
\usepackage{ulem}
\usepackage{xcolor}
\usepackage{xr-hyper}
\makeatother

\begin{document}
\title{Tuning electronic properties and Schottky contact in graphene-based van der Waals heterostructures by electric gating and interlayer coupling}
\author{Poonam Sharma}
\thanks{These authors contributed equally to this work.}
\author{Archana Sharma}
\thanks{These authors contributed equally to this work.}
\author{Alok Shukla}
\email{shukla@iitb.ac.in}

\affiliation{Department of Physics, Indian Institute of Technology Bombay, Mumbai
400076, India}
\begin{abstract}
Van der Waals heterostructures incorporating graphene have been an active area of research, both theoretical and experimental, due to their potential to yield devices with a wide variety of applications. In this paper, first-principles calculations are employed to investigate C$_{6}$N$_{6}$/graphene, hg-C$_{3}$N$_{4}$/graphene, and C$_{6}$N$_{6}$/hg-C$_{3}$N$_{4}$ two-dimensional van der Waals heterostructures. A systematic analysis of structural and thermodynamic stability, electronic, mechanical, and optical properties of semiconductor/metal (C$_{6}$N$_{6}$/graphene and hg-C$_{3}$N$_{4}$/graphene) and semiconductor/semiconductor (C$_{6}$N$_{6}$/hg-C$_{3}$N$_{4}$) interfaces is performed. Both semiconductor/metal heterostructures form $n$-type Schottky contacts, which can be converted into $p$-type Schottky or Ohmic contacts by tuning the external perpendicular electric field and the interlayer coupling. In the semiconductor/semiconductor C$_{6}$N$_{6}$/hg-C$_{3}$N$_{4}$ heterostructure, the valence and conduction band edges originate from distinct layers, resulting in a type-II band alignment that promotes efficient electron-hole separation. Furthermore, the band alignment can be effectively tuned between type-I and type-II by applying an external electric field and varying the interlayer distance. From the optical absorption spectra of the heterostructures, we concluded that the C$_{6}$N$_{6}$/graphene and hg-C$_{3}$N$_{4}$/graphene heterostructures exhibit an optical response across a wide frequency range, whereas the C$_{6}$N$_{6}$/hg-C$_{3}$N$_{4}$ heterostructure shows prominent activity primarily in the ultraviolet region. Using the $G_{0}W_{0}$+BSE approach, the exciton binding energies are also calculated for the gapped systems, namely C$_{6}$N$_{6}$, hg-C$_{3}$N$_{4}$ monolayers, and their heterostructure (C$_{6}$N$_{6}$/hg-C$_{3}$N$_{4}$), yielding values of 1.01~eV, 1.14~eV, and 1.18~eV, respectively, highlighting strong electron-hole (e-h) interactions. Moreover, the band-edge analysis of C$_{6}$N$_{6}$/hg-C$_{3}$N$_{4}$ heterostructure further favors pronounced interlayer e-h coupling. These findings provide valuable insights into the design and tuning of the two-dimensional van der Waals heterostructures, paving the way for advanced applications in nanoelectronics, optoelectronics, and photocatalysis.
 
\end{abstract}
\maketitle

\section{Introduction}

\label{sec:level1}

The successful discovery of graphene (GE) in 2004 by Geim \textit{et al.} through
graphite exfoliation marked an important milestone in the study of
two-dimensional (2D) materials~\cite{doi:10.1126/science.1102896}.
Since then, significant efforts have been made to investigate the
fascinating properties, applications, and synthesis techniques of
GE, leading to new research avenues. The GE monolayer consisting of
carbon (C) atoms with the sp$^{2}$-hybridized arrangement in a honeycomb
pattern exhibits peculiar and intriguing optoelectronic, thermal,
and mechanical properties~\cite{RevModPhys.81.109,PhysRevB.85.125428}.
Along with high carrier mobility~\cite{PhysRevLett.100.016602},
GE exhibits many other remarkable properties such as unusual quantum
hall effect~\cite{PhysRevLett.124.136403}, high thermal conductivity~\cite{doi:10.1021/nl0731872},
indicating its possible applications in a variety of fields such as
batteries~\cite{doi:10.1021/nl200658a}, sensors~\cite{PhysRevApplied.22.014071},
and transport applications~\cite{RevModPhys.82.2673}. While having
all these exceptional qualities, the zero band gap of GE limits its
application in modern technology. For example, in transistors, the
very low on-off ratio and excessively large off-state current restrict
their use in nano-electronics applications~\cite{doi:10.1021/nl9039636}.
Additionally, the restacking phenomenon of GE nanosheets significantly
reduces the gas detection capability of its thin films and poses a
considerable obstacle to the production of electrodes for energy storage
devices. These challenges have influenced and driven scientific research
into the development of novel 2D materials over the past decade.

Consequently, a large number of 2D materials, such as silicene, transition-metal
dichalcogenides, and phosphorene have been investigated for a wide
range of technological applications~\cite{vogt2012silicene,sharma2024defect,PhysRevLett.114.066803}.
In low-dimensional materials, the size and geometry primarily determine
their optoelectronic characteristics. Moreover, perforation of 2D
Materials have also been found to enhance their optoelectronic properties,
thereby broadening their scope of applications and hence gaining immense
interest lately. For instance, holey graphene (HG), namely GE with
periodic tiny pores (holes), exhibits semiconducting behavior and
shows promising applications in energy storage, optoelectronics, and
thermoelectrics. It is also found that the electronic band gap in
these types of systems depends on the size of the pores. The size
of the pores allows for tuning of the electronic band gap. As a result,
various GE-based structures have been explored for different applications
by varying pore sizes~\cite{PhysRevB.92.085421,PhysRevLett.108.086804}.

Moreover, in addition to exploring new materials, significant research
has been devoted to opening the band gap in GE. This has involved
various techniques, such as applying strain~\cite{PhysRevB.78.075435},
substrate effects~\cite{PhysRevB.76.073103}, intercalating with
different monolayers~\cite{PhysRevB.100.115439}, and substitutional
doping~\cite{PhysRevB.83.155445}. Such modifications present new
opportunities for tuning the band gap and managing the spin polarization
characteristics of GE. In substitutional doping, boron (B) and nitrogen
(N) substitutions are mainly studied from both experimental and theoretical
perspectives because their atomic radii are almost the same as that
of the C atom~\cite{doi:10.1126/science.1208759,PhysRevB.90.125418,PhysRevB.84.125401}.
Further, introducing nitrogen into the GE lattice serves as a method
to widen the band gap, thereby enhancing the performance of electronic
devices. As a result, a series of new 2D materials, known as carbon
nitrides C$_{n}$N$_{m}$ (where n and m denote the number of carbon
(C) and nitrogen atoms, respectively), have also been investigated
by combining the C with N atoms in different stoichiometries~\cite{sakhraoui2023}.
These materials, which bear a geometrical resemblance to GE, exhibit
remarkable physical and chemical properties.\\
Further, forming the metal/semiconductor interface contact is also an efficient
way to construct high-efficiency and energy-conserving electronic
devices. Achieving minimal or negligible contact resistance at a metal-semiconductor
junction is crucial for the development of high-performance devices
that rely on Ohmic contacts. Factors like the mismatch in work function,
surface imperfections, and difficulties in applying reliable doping
methods make it particularly hard to achieve Ohmic contact at metal-2D
semiconductor interfaces. These interfaces often exhibit Schottky
characteristics, complicating the process of realizing Ohmic
behavior. Considerable research has focused on achieving quasi-Ohmic
contacts to remove or reduce the Schottky barrier height (SBH). Applying
a high gate voltage can shift a metal-semiconductor system from the
Ohmic to Schottky regime in certain cases. For effective device performance,
it is essential to establish a natural Ohmic contact at the metal-semiconductor
interface, which remains challenging both experimentally and theoretically.
A key method to reduce the SBH involves the use of strain and electric
field modulation techniques, which further preserve the intrinsic
properties of the materials involved.\\
 In the present work, a systematic investigation of two C$_{n}$N$_{m}$
monolayers, i.e., C$_{6}$N$_{6}$ and hg-C$_{3}$N$_{4}$, is performed
using density functional theory (DFT) calculations. Along with studying
their structural, electronic, magnetic, and optical properties, we
combined each monolayer with a GE monolayer to construct van der Waals
(vdW) heterostructures (HTSs) in order to explore their potential
as metal-semiconductor contacts. Both C$_{6}$N$_{6}$/GE and hg-C$_{3}$N$_{4}$/GE
HTSs are found to form $n$-type Schottky contacts ($n$-SCs). Additionally, we
constructed a C$_{6}$N$_{6}$/hg-C$_{3}$N$_{4}$ HTS, which does
not exhibit a metal-semiconductor contact but allows further insight
into the interaction between different C$_{n}$N$_{m}$ layers. The
effects of the perpendicular electric field and interlayer coupling
on the electronic properties of all considered 2D vdW HTSs are also
analyzed.

\section{{COMPUTATIONAL DETAILS}}

\label{sec:level1}

First-principles density functional theory (DFT) calculations were
performed within the Perdew-Burke-Ernzerhof exchange-correlation functional
of generalized gradient approximation (GGA) using the Vienna~\textit{ab initio} simulation package (\textsc{vasp})~\cite{PhysRevLett.77.3865,PhysRevB.54.11169}.
We used the projector augmented wave pseudopotentials for both carbon
and nitrogen atoms with the electronic configurations s$^{2}$p$^{2}$
and s$^{2}$p$^{3}$, respectively~\cite{PhysRevB.50.17953,PhysRevB.59.1758}.
For geometry optimization, a kinetic energy cutoff of 500 eV was used,
with a convergence criterion of 1$\times$10$^{-2}$~eV/\AA~for
the Hellmann-Feynman forces on each atom. For the Brillouin zone sampling, a $\Gamma$-centred k-mesh of 23$\times$23$\times$1 was used for all considered unit cell structures, while a 12$\times$12$\times$1 $\Gamma$-centred k-mesh was employed for the HTSs.
To prevent interactions between neighboring layers, a substantial, large vacuum of 21~\AA~is
considered in the calculations. In order to account for the vdW interactions
between layers of HTSs, Grimme's DFT-D3 empirical correction is incorporated
in our calculations~\cite{doi:10.1021/jp106469x,doi:10.1021/acs.nanolett.9b02982}.
We also repeated some of our calculations using the HSE06 functional
to verify our GGA-based results. For the investigation of the optical
properties of the considered HTSs and isolated monolayers, the random-phase
approximation (RPA) was employed. Within this formalism, the linear
density response to an external perturbing electric field is described
in the independent-particle picture, thereby neglecting excitonic
effects. The macroscopic dielectric function $\varepsilon(\omega)$
is obtained from the independent-particle polarizability constructed
using the single-particle orbitals and eigenvalues derived from the
preceding GGA-DFT calculations. A $12\times12\times1$ $\Gamma$-centered
k-point mesh was used for the monolayers, while a $6\times6\times1$ $\Gamma$-centered mesh was adopted for the HTSs. From the calculated
macroscopic dielectric function, the imaginary part of the dielectric
constant, $\varepsilon_{2}(\omega)$, arising from interband optical
transitions, is evaluated using the momentum matrix elements between
occupied (valence) and unoccupied (conduction) states as

\begin{equation}
\varepsilon_{2}(\omega)=\frac{4\pi^{2}e^{2}}{\Omega}\sum_{\mathbf{k},v,c}w_{\mathbf{k}}\left|\left\langle \psi_{c\mathbf{k}}\middle|\mathbf{u}\cdot\mathbf{p}\middle|\psi_{v\mathbf{k}}\right\rangle \right|^{2}\delta(E_{c\mathbf{k}}-E_{v\mathbf{k}}-\hbar\omega),\label{eq:eps2}
\end{equation}

where $\Omega$ is the unit-cell volume, $w_{\mathbf{k}}$ is the
k-point weight, $\psi_{n\mathbf{k}}$ are the Bloch wave functions,
$\mathbf{u}$ is the polarization direction of the incident electric
field, and $\mathbf{p}$ is the momentum operator. The real part of
the dielectric constant, $\varepsilon_{1}(\omega)$, is then obtained
from $\varepsilon_{2}(\omega)$ through the Kramers-Kronig relation
as 
\begin{equation}
\varepsilon_{1}(\omega)=1+\frac{2}{\pi}\,\mathcal{P}\int_{0}^{\infty}\frac{\omega'\varepsilon_{2}(\omega')}{\omega'{}^{2}-\omega^{2}}\,d\omega',\label{eq:eps1}
\end{equation}
where $\mathcal{P}$ denotes the principal value of the integral.
Once the real and imaginary parts of the dielectric function are obtained,
various frequency-dependent optical quantities such as the refractive
index $n(\omega)$, extinction coefficient $k(\omega)$, absorption
coefficient $\alpha(\omega)$, reflectivity $R(\omega)$, and electron
energy-loss function $L(\omega)$ are computed using

\begin{align}
n(\omega) & =\frac{1}{\sqrt{2}}\left[\left(\varepsilon_{1}^{2}(\omega)+\varepsilon_{2}^{2}(\omega)\right)^{1/2}+\varepsilon_{1}(\omega)\right]^{1/2},\\
k(\omega) & =\frac{1}{\sqrt{2}}\left[\left(\varepsilon_{1}^{2}(\omega)+\varepsilon_{2}^{2}(\omega)\right)^{1/2}-\varepsilon_{1}(\omega)\right]^{1/2},\\
\alpha(\omega) & =\frac{2\omega\,k(\omega)}{c},\\
R(\omega) & =\left|\frac{n(\omega)-1+ik(\omega)}{n(\omega)+1+ik(\omega)}\right|^{2},\\
L(\omega) & =\mathrm{Im}\left[-\frac{1}{\varepsilon(\omega)}\right].
\end{align}

Further, we also calculated the exciton binding
energies for the considered systems using the G$_{0}$W$_{0}$ approach
combined with the Bethe-Salpeter equation (BSE). Our discussion of
exciton binding energy is limited only to gapped systems, as the concept
of a bound exciton and its binding energy in the conventional G$_{0}$W$_{0}$+BSE
formalism is not meaningful for metallic systems. The
BSE was solved using the single-shot G$_{0}$W$_{0}$ quasiparticle
energies. The G$_{0}$W$_{0}$+BSE approach accounted for electron-electron
and electron-hole interactions. During the calculations, we have used
the same k-point sampling as used in the RPA, namely 12$\times$12$\times$1
$\Gamma$-centered k-point sampling for the monolayers and 6$\times$6$\times$1
$\Gamma$-centered k-point sampling for the HTSs. The plane-wave cutoff
energy was set to 400 eV, while the cutoff for the response function
was fixed at 260 eV. The quasiparticle energies obtained from the
preceding G$_{0}$W$_{0}$ calculations were used as input for the
BSE calculations. For each system, the total number of bands included
in the ground-state calculations (NBANDS) and the number of occupied
bands used to construct the electron-hole excitation space in the
BSE calculations (NBANDSO) were carefully converged. The number of
unoccupied bands (NBANDSV) was set to twice NBANDSO to provide a sufficiently
large virtual space for obtaining converged optical spectra. The values
of NBANDSO employed for the studied gapped systems were 27 (C$_{6}$N$_{6}$),
32 (hg-C$_{3}$N$_{4}$), and 59 (C$_{6}$N$_{6}$/hg-C$_{3}$N$_{4}$).
The solution of the BSE Hamiltonian yields the excitonic states, from which the macroscopic frequency-dependent dielectric function is obtained. From the complex dielectric function, we computed the absorption coefficient
$\alpha(\omega)$, to determine the optical band gap. Further,
the exciton binding energy ($E_{b}^{exc}$) is calculated as the difference
between the fundamental quasiparticle band gap and the optical gap
\begin{equation}
E_{b}^{exc}=E_{g}^{\mathrm{QP}}-E_{g}^{\mathrm{opt}},\label{excit}
\end{equation}
where $E_{g}^{\mathrm{QP}}$ is the quasiparticle band gap obtained from single-shot $G_{0}W_{0}$ calculations and $E_{g}^{\mathrm{opt}}$ is the optical gap obtained from the solution of the BSE. The quasiparticle band gap $E_{g}^{\mathrm{QP}}$ is determined from the $G_{0}W_{0}$ quasiparticle eigenvalues. To obtain a well-resolved quasiparticle band dispersion along the high-symmetry paths, these eigenvalues were interpolated using maximally localized Wannier functions as implemented in Wannier90~\cite{mostofi2008wannier90}. The optical gap $E_{g}^{\mathrm{opt}}$ corresponds to the lowest-energy excitation obtained from the BSE spectrum.

\section{RESULTS AND DISCUSSION}

\subsection{Structural and electronic properties of 2D monolayers}

\begin{figure}[ht]
\includegraphics[width=1\linewidth]{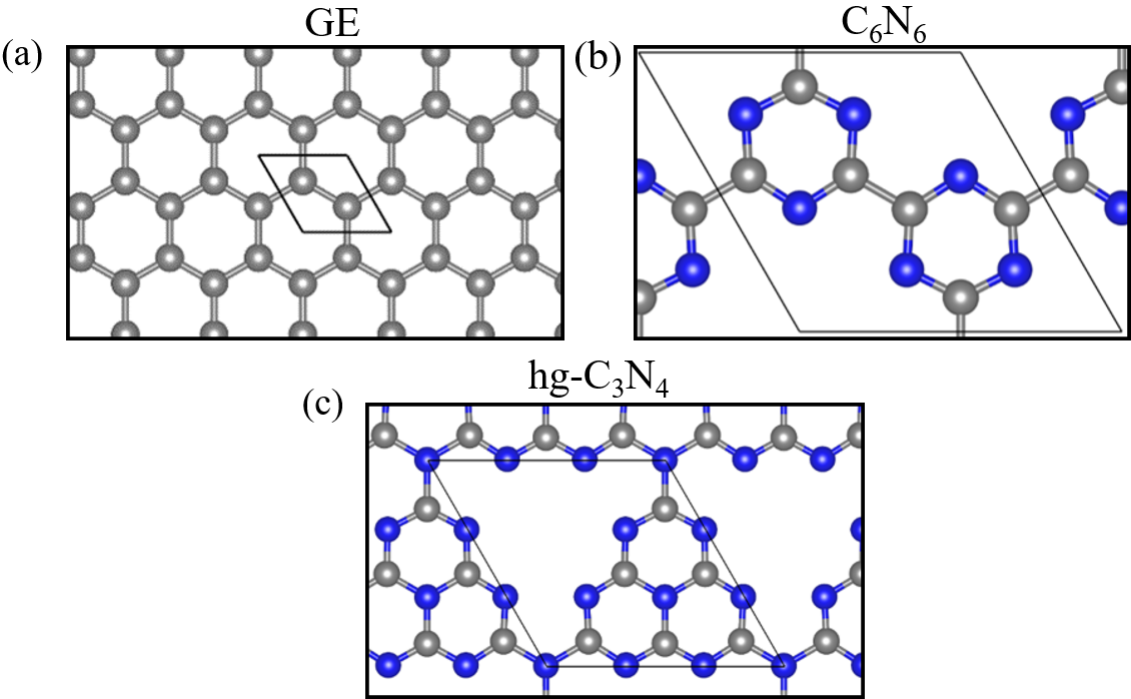} \caption{Optimized structures of (a) GE, (b) C$_{6}$N$_{6}$, and (c) hg-C$_{3}$N$_{4}$
monolayers. The black parallelogram inside each structure indicates
the unit cell of the respective structure. Gray and blue colors represent
C and N atoms, respectively.}
\label{fig:pris_str} 
\end{figure}

First, we have examined the structure and electronic properties of
different monolayers considered in this study. Figs.~\ref{fig:pris_str}(a)-\ref{fig:pris_str}(c)
show the optimized structures corresponding to the GE, C$_{6}$N$_{6}$
and hg-C$_{3}$N$_{4}$ monolayers.

\begin{table*}[ht]
\caption{Calculated optimized lattice parameter ($a$), minimum C-C ($d_{\text{C-C}}$)
and C-N ($d_{\text{C-N}}$) bond lengths, work function (W), band
gap (E$_{g}$) values, and the nature of the band gaps for GE, C$_{6}$N$_{6}$,
and hg-C$_{3}$N$_{4}$ monolayers using the PBE functional. Band
gap values for C$_{6}$N$_{6}$ and hg-C$_{3}$N$_{4}$ were also
obtained using the HSE06 functional.}

\begin{ruledtabular}
\begin{tabular}{c|c|c|c|c|c|c|cc}
Monolayer  & a ({\AA})  & $d$$_{C-C}$ ({\AA})  & $d$$_{C-N}$ ({\AA})  & E$_{g}$ PBE (eV)  & E$_{g}$ HSE06 (eV)  & nature  & W PBE/HSE06 (eV)  & \tabularnewline
\hline 
GE  & 2.46  & 1.42  & ...  & 0  & ...  & SM at $\Gamma$  & 4.26/4.51  & \tabularnewline
C$_{6}$N$_{6}$  & 7.11  & 1.51  & 1.34  & 1.53  & 3.18  & Direct at K  & 5.78/7.49  & \tabularnewline
hg-C$_{3}$N$_{4}$  & 7.13  & ...  & 1.33  & 1.18  & 2.98  & Indirect $\Gamma$$\rightarrow$ K  & 4.63/6.56  & \tabularnewline
\end{tabular}\label{tab:parameter} 
\end{ruledtabular}

\end{table*}

\begin{figure}[ht]
\includegraphics[width=1\linewidth]{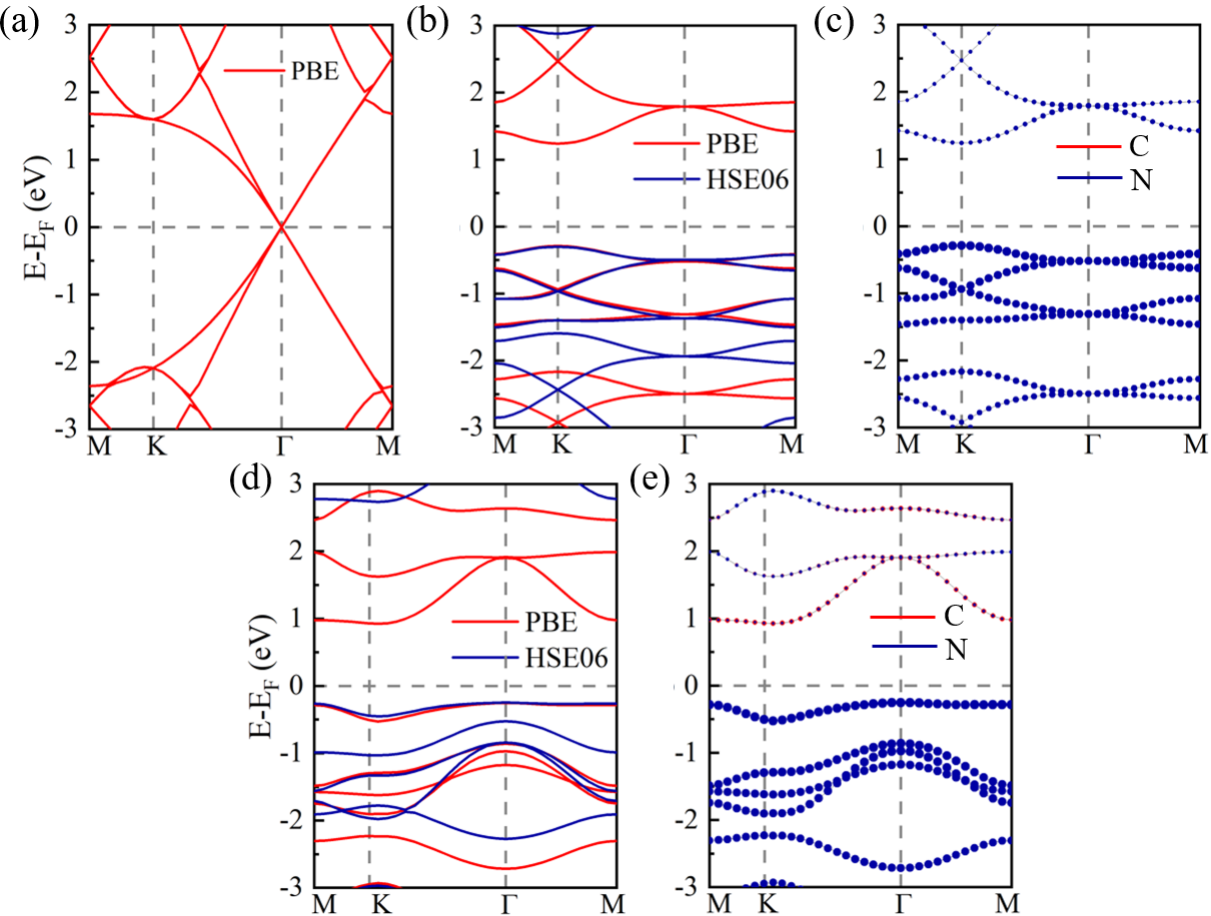} \caption{Electronic band structures of (a) GE (using PBE only), (b) C$_{6}$N$_{6}$ monolayer (showing both PBE and HSE06 results), and (d) hg-C$_{3}$N$_{4}$ monolayer (showing both PBE and HSE06 results). (c), (e) Atom-projected band structures of C$_{6}$N$_{6}$ and hg-C$_{3}$N$_{4}$ monolayers, respectively, obtained using PBE.}
\label{fig:bandpris} 
\end{figure}

The optimized lattice parameters, C-C bond lengths, C-N bond lengths,
band gaps, the nature of the band gaps, and the work functions for these planar monolayer structures are listed in Table~\ref{tab:parameter}, which agree well with the previously reported results~\cite{bafekry2019two,PhysRevB.102.134112}.
Figs.~\ref{fig:bandpris}(a)-\ref{fig:bandpris}(e) show the band structure plots for all considered monolayers. The GE monolayer is a well-known Dirac-semimetal (SM) with zero gap (See Fig.~\ref{fig:bandpris}(a)). Moreover, band structure calculations performed using the PBE/HSE06 methods reveal that C$_{6}$N$_{6}$ is a direct band gap, non-magnetic semiconductor with a band gap of 1.53/3.18 eV at the K point (See Fig.~\ref{fig:bandpris}(b)), whereas hg-C$_{3}$N$_{4}$ is an indirect band gap, non-magnetic semiconductor with a band gap of 1.18/2.98
eV, spanning from the valence band maximum (VBM) at the $\Gamma$
point to the conduction band minimum (CBM) at the K point (See Fig.~\ref{fig:bandpris}(d)). These obtained
band gaps are in good agreement with the previously reported values~\cite{bafekry2019two,PhysRevB.102.134112}. It is important to note that for a two-atom primitive cell, the Dirac point of the GE monolayer is at the K point. However, because in our calculations a 3 \texttimes 3
\texttimes 1 supercell is used, it gets folded to the $\Gamma$ point, and hence appears at the center of the Brillouin zone. Further, from the projected band structures of C$_{6}$N$_{6}$ (see Fig.~\ref{fig:bandpris}(c)) and hg-C$_{3}$N$_{4}$ (see Fig.~\ref{fig:bandpris}(e)) monolayers, it was found that in the case C$_{6}$N$_{6}$ monolayer, both in CBM and VBM, N atoms contribute mainly, whereas, in the hg-C$_{3}$N$_{4}$ monolayer, N atoms contribute mainly to the VBM, whereas in the CBM, the maximum contribution is obtained from the C atoms.

\subsection{C$_{6}$N$_{6}$/GE, hg-C$_{3}$N$_{4}$/GE, and C$_{6}$N$_{6}$/hg-C$_{3}$N$_{4}$
van der Waals heterostructures}

\subsubsection{Structural analysis and stability}
To construct the interface between GE and the considered monolayers, we considered bilayers composed of C$_{6}$N$_{6}$ and hg-C$_{3}$N$_{4}$ monolayers, on top of the GE monolayer. This arrangement forms two distinct HTSs, namely C$_{6}$N$_{6}$/GE and hg-C$_{3}$N$_{4}$/GE. A 3$\times$3$\times$1 GE supercell was used to minimize lattice mismatch during construction of the HTSs. Additionally, a mixed HTS between C$_{6}$N$_{6}$ and hg-C$_{3}$N$_{4}$, referred to as C$_{6}$N$_{6}$/hg-C$_{3}$N$_{4}$,
where hg-C$_{3}$N$_{4}$ is the lower layer, is also considered. The lattice mismatch ($\Delta$$\lambda$) between the two monolayers
was calculated using the following formula 
\begin{equation}
\Delta\lambda=\frac{|a_{top}-a_{bottom}|}{|a_{bottom}|}\times100\%,
\end{equation}
where $a_{top}$ and $a_{bottom}$ are the lattice parameters of the
top and bottom layers, respectively. This approach optimizes the interfaces
in the HTS, thereby enhancing the potential performance of the resulting
configurations. By considering a 3$\times$3$\times$1 GE supercell,
the lattice mismatch was reduced to 3.66\% between GE and C$_{6}$N$_{6}$
and to 3.39\% between GE and hg-C$_{3}$N$_{4}$ HTSs. In the case
of C$_{6}$N$_{6}$/hg-C$_{3}$N$_{4}$ HTS, a lattice mismatch of
less than 1$\%$ is noticed. The minimal lattice distortion (i.e.,
$<$ 5\%) within these HTSs enables accurate \textit{ab initio} calculations.
\begin{figure*}[ht]
\includegraphics[width=1\linewidth]{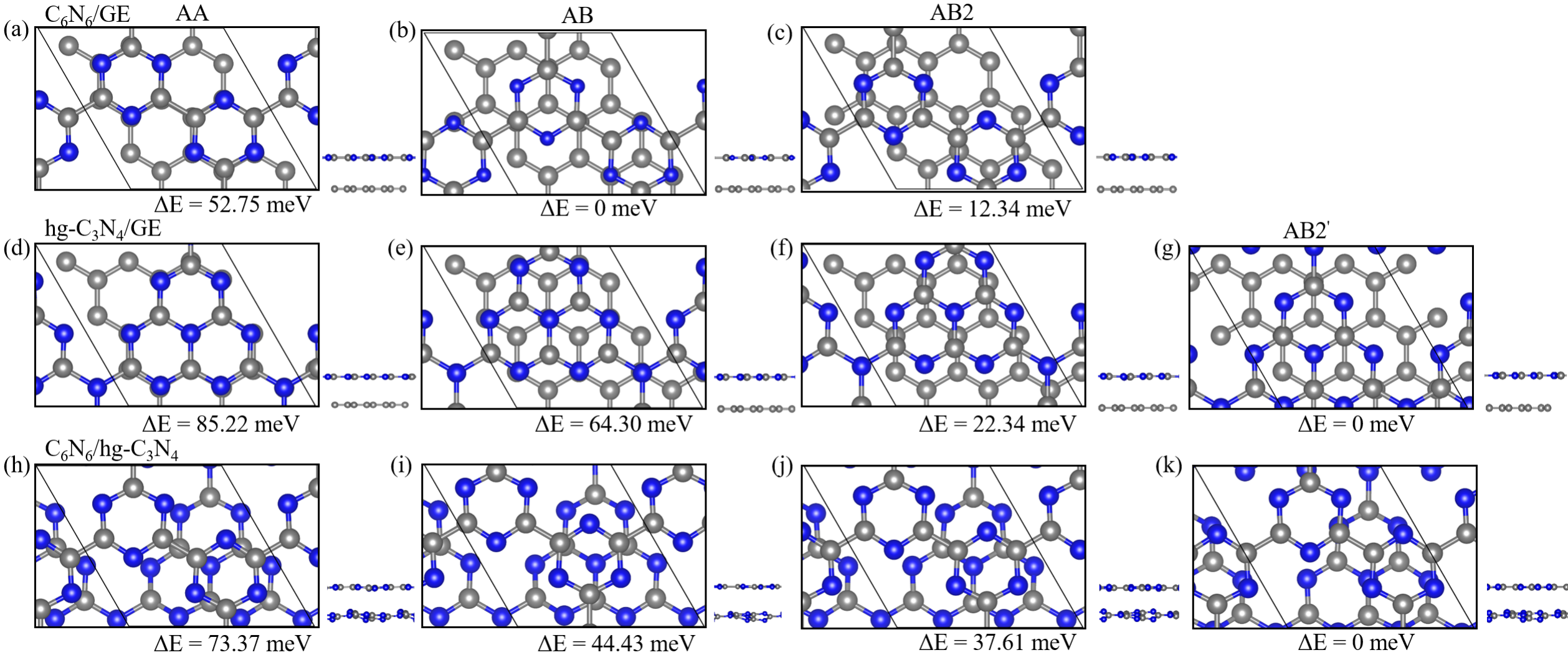}\caption{Top and side views of different possible stacking configurations of
(a)-(c) C$_{6}$N$_{6}$/GE, (d)-(g) hg-C$_{3}$N$_{4}$/GE, and (h)-(k)
C$_{6}$N$_{6}$/hg-C$_{3}$N$_{4}$ HTSs. The $\Delta$E represents
the difference in energy with respect to the most stable configuration.}
\label{diff_HTSs} 
\end{figure*}

\begin{table*}[ht]
\caption{Computed lattice parameter ($a$), minimum C-C ($d_{\text{C-C}}$)
and C-N ($d_{\text{C-N}}$) bond lengths, interlayer distance ($d_{\text{inter}}$),
buckling ($\delta$), band gap ($\mathrm{E}_{\mathrm{g}}$) values,
and work function (W) for the vdW heterostructures (HTSs) C$_{6}$N$_{6}$/GE,
hg-C$_{3}$N$_{4}$/GE, and C$_{6}$N$_{6}$/hg-C$_{3}$N$_{4}$ using the
PBE functional. Band gap values for all three HTSs were also obtained
using the HSE06 functional. While reporting bond lengths (i.e., $d_{\mathrm{C-C}}$
and $d_{\mathrm{C-N}}$), the first and second values correspond to
the A and B layers, respectively, following the heterostructure notation
A/B. NA indicates that the corresponding C-C or C-N bond does not
exist in the corresponding monolayer. }

\begin{ruledtabular}
\begin{tabular}{c|c|c|c|c|c|c|c|cc}
HTS  & a ({\AA})  & d$_{CC}$ ({\AA})  & d$_{NC}$ ({\AA})  & d$_{inter}$ ({\AA})  & $\delta$ ({\AA})  & $\mathrm{E}_{\mathrm{g}}$ (PBE/HSE06)  & nature  & W (eV)  & \tabularnewline
\hline 
C$_{6}$N$_{6}$/GE  & 7.33  & 1.61/1.42  & 1.36/NA  & 3.38  & 0,0  & 39.70/30.60 meV  & Direct at $\Gamma$  & 4.55  & \tabularnewline
hg-C$_{3}$N$_{4}$/GE  & 7.40  & NA/1.42  & 1.36/NA  & 3.37  & 0,0  & 25.50/27.80 meV  & Direct at $\Gamma$  & 4.38  & \tabularnewline
C$_{6}$N$_{6}$/hg-C$_{3}$N$_{4}$  & 7.13  & NA/1.51  & 1.38/1.34  & 2.98  & 0.88/0  & 1.12/2.59 eV  & Indirect $\Gamma$$\rightarrow$ K/Direct at K  & 5.40  & \tabularnewline
\end{tabular}\label{tab:parofhts} 
\end{ruledtabular}

\end{table*}

In each HTS, various possible stacking configurations are considered
as shown in Fig.~\ref{diff_HTSs}, and the most stable configuration,
corresponding to the lowest energy state, is indicated by $\Delta$E
= 0. For C$_{6}$N$_{6}$/GE, the AB stacking configuration is found
to be the most stable, whereas hg-C$_{3}$N$_{4}$/GE and C$_{6}$N$_{6}$/hg-C$_{3}$N$_{4}$
HTSs are found to be most stable in the AB2' stacking configuration.
The optimized lattice parameter ($a$), minimum C-C ($d_{\text{C-C}}$)
and C-N ($d_{\text{C-N}}$) bond lengths, interlayer distance (d$_{inter}$), buckling
parameter ($\delta$), and the work function ($W$) of the 2D vdW
HTSs corresponding to the most stable configuration are presented
in Table~\ref{tab:parofhts}. The obtained equilibrium interlayer
distances in different considered HTSs are within the range of other
studied vdW HTSs, indicating that with such values of interlayer distances,
these HTSs are characterized by weak vdW forces~\cite{PhysRevLett.114.066803,PhysRevB.97.235419}.
Next, we computed the interface binding energy (E$_{b}$) of the considered
HTSs corresponding to the most stable configurations using the formula
E$_{b}$ = (E$_{C_{n}N_{m}/X}$ - E$_{C_{n}N_{m}}$ - E$_{X}$)/A,
where E$_{C_{n}N_{m}/X}$ is the total energy of the HTS, E$_{C_{n}N_{m}}$
is the total energy of the C$_{n}$N$_{m}$ monolayer, E$_{X}$ is
the total energy of the other monolayer, and A is the surface area
of the HTS interface at equilibrium. In C$_{6}$N$_{6}$/GE HTS, the
binding energy of -19 meV/\AA$^{2}$, while in hg-C$_{3}$N$_{4}$/GE
and C$_{6}$N$_{6}$/hg-C$_{3}$N$_{4}$ HTSs, the binding energies
of -21 meV/\AA$^{2}$ and -36 meV/\AA$^{2}$ are obtained, respectively.
The obtained negative value of E$_{b}$ in each case shows the structural
stability of all considered HTSs. Moreover, these obtained E$_{b}$
values are comparable with the other 2D vdW HTSs~\cite{PhysRevB.92.195419,PhysRevB.97.235419,PhysRevB.100.115439},
which further confirms the C$_{n}$N$_{m}$ monolayer is bounded with
GE or to another C$_{n}$N$_{m}$ monolayer through weak vdW forces
and shows the possibility of experimental synthesis of the considered
HTSs. Further, to examine the thermal stability of the considered
HTSs, we performed \textit{ab initio} molecular dynamics (AIMD) simulations
for a period of 5 ps at room temperature (300 K). The variation in total energy as a function of time step is shown in Figs. S1(a)-S1(c) of the Supplemental Material (SM)~\cite{SI}. After
5 ps, the structures remain stable in all considered HTSs, and the fluctuation in total energy over time is minimal. These results demonstrate that the considered HTSs are stable at room temperature, making them
promising candidates for practical applications.

\subsubsection{Mechanical properties}

To assess the mechanical stability and in-plane stiffness of the C$_{6}$N$_{6}$/GE,
hg-C$_{3}$N$_{4}$/GE, and C$_{6}$N$_{6}$/hg-C$_{3}$N$_{4}$ HTSs,
the in-plane elastic constants ($C_{11}$, $C_{12}$, and $C_{66}$)
were evaluated using the energy-strain method and presented in Table~\ref{Table:elas}
( See Fig. S2 of the SM~\cite{SI}). In all cases, the calculated elastic constants
satisfy the Born-Huang mechanical stability criteria~\cite{PhysRevB.90.224104},
namely $C_{11}>|C_{12}|$ and $C_{66}=\frac{1}{2}(C_{11}-C_{12})>0$,
signifying mechanical stability and indicating an isotropic in-plane
elastic response, as expected for systems with threefold rotational
symmetry. Moreover, the corresponding Young’s modulus ($E(\theta)$)
and Poisson’s ratio ($\nu(\theta)$), also listed in Table~\ref{Table:elas},
show that the C$_{6}$N$_{6}$/GE and hg-C$_{3}$N$_{4}$/GE HTSs
possess higher stiffness and smaller $\nu(\theta)$ compared to pristine
graphene, while the C$_{6}$N$_{6}$/hg-C$_{3}$N$_{4}$ HTS exhibits
relatively lower rigidity. Overall, the computed elastic parameters
confirm that all HTSs are mechanically stable and exhibit isotropic
in-plane mechanical properties, indicating their suitability for flexible
and durable 2D material applications.

\begin{table}[ht]
\centering 
\caption{ Elastic properties, including in-plane elastic stiffness constants,
Young’s modulus ($E(\theta)$) (units: N/m), and Poisson’s ratio ($\nu(\theta)$),
for the C$_{6}$N$_{6}$/GE, hg-C$_{3}$N$_{4}$/GE, and C$_{6}$N$_{6}$/hg-C$_{3}$N$_{4}$
heterostructures, calculated using GGA.}
\label{tab:table1} 
\begin{ruledtabular}
\begin{tabular}{cccccc}
System  & $C_{11}$  & $C_{12}$  & $C_{66}$  & $E(\theta)$  & $\nu(\theta)$\tabularnewline
\hline 
C$_{6}$N$_{6}$/GE  & 467.25  & 23.33  & 221.96  & 466.08  & 0.05\tabularnewline
hg-C$_{3}$N$_{4}$/GE  & 435.90  & 34.01  & 200.95  & 433.25  & 0.08\tabularnewline
C$_{6}$N$_{6}$/hg-C$_{3}$N$_{4}$  & 189.19  & 12.48  & 88.35  & 188.37  & 0.07\tabularnewline
GE~\cite{PhysRevB.82.235414,PhysRevB.105.235303}  & 349  & 72  & 138  & 334  & 0.21\tabularnewline
\end{tabular}
\end{ruledtabular}

\label{Table:elas} 
\end{table}

\subsubsection{Electronic properties and charge transfer Characteristics}

\begin{figure}[ht]
\includegraphics[width=1\linewidth]{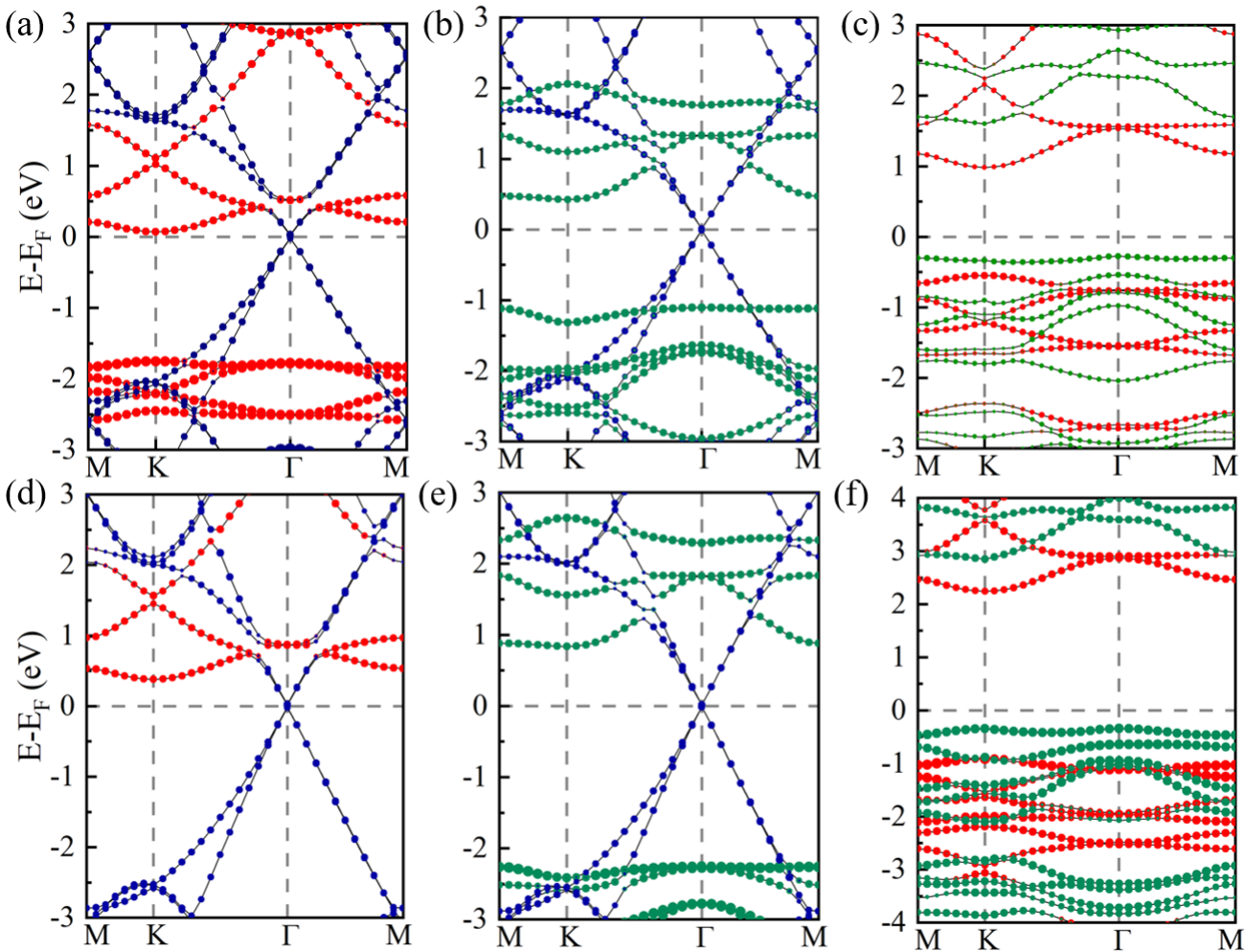}\caption{Layer-projected band structures of (a), (d) C$_{6}$N$_{6}$/GE, (b), (e)
hg-C$_{3}$N$_{4}$/GE, and (c), (f) C$_{6}$N$_{6}$/hg-C$_{3}$N$_{4}$
heterostructures, computed using PBE (top panels: (a)-(c)) and HSE06
(bottom panels: (d)-(f)) functionals. In all band structure plots,
the blue color represents the contribution of the GE monolayer within
the respective heterostructure, while the red and green colors indicate
the contributions of the C$_{6}$N$_{6}$ and hg-C$_{3}$N$_{4}$
monolayers, respectively. The Fermi level is set to zero eV in all
band structure plots.}
\label{fig:EBS} 
\end{figure}

Next, we investigated the electronic properties of the considered HTSs. Figs.~\ref{fig:EBS}(a)-\ref{fig:EBS}(f) present the layer-projected band structures of C$_{6}$N$_{6}$/GE, hg-C$_{3}$N$_{4}$/GE, and
C$_{6}$N$_{6}$/hg-C$_{3}$N$_{4}$ HTSs, computed using the PBE
(top panels) and HSE06 (bottom panels) methods, respectively. The obtained band structures of the considered HTSs reflect the combined electronic characteristics of the individual monolayers. The HTSs involving GE continue to exhibit Dirac cones near the Fermi energy, but with small gap openings of 45.68/30.6 meV for the C$_{6}$N$_{6}$/GE
HTS and 21.91/27.8 meV for the hg-C$_{3}$N$_{4}$/GE HTS, as computed
using the PBE/HSE06 method. In the case of C$_{6}$N$_{6}$/hg-C$_{3}$N$_{4}$
HTS, a band gap of 1.12/2.59 eV is obtained using the PBE/HSE06 method~\cite{liang2016photocatalytic}.
The obtained band gaps corresponding to all three HTSs using PBE and
HSE06 methods are reported in Table~\ref{tab:parofhts}. Further, the layer-projected band structures of C$_{6}$N$_{6}$/GE and hg-C$_{3}$N$_{4}$/GE
HTSs also reveal that close to the Fermi level, where the dispersion
is linear, the bands are dominated by the contribution of the GE orbitals.
However, as we move away from the Fermi energy, the dispersion ceases
to be linear, and the bands exhibit strong hybridization with the
semiconducting layer's orbital contributions dominating.

\begin{figure}[ht]
\includegraphics[width=1\linewidth]{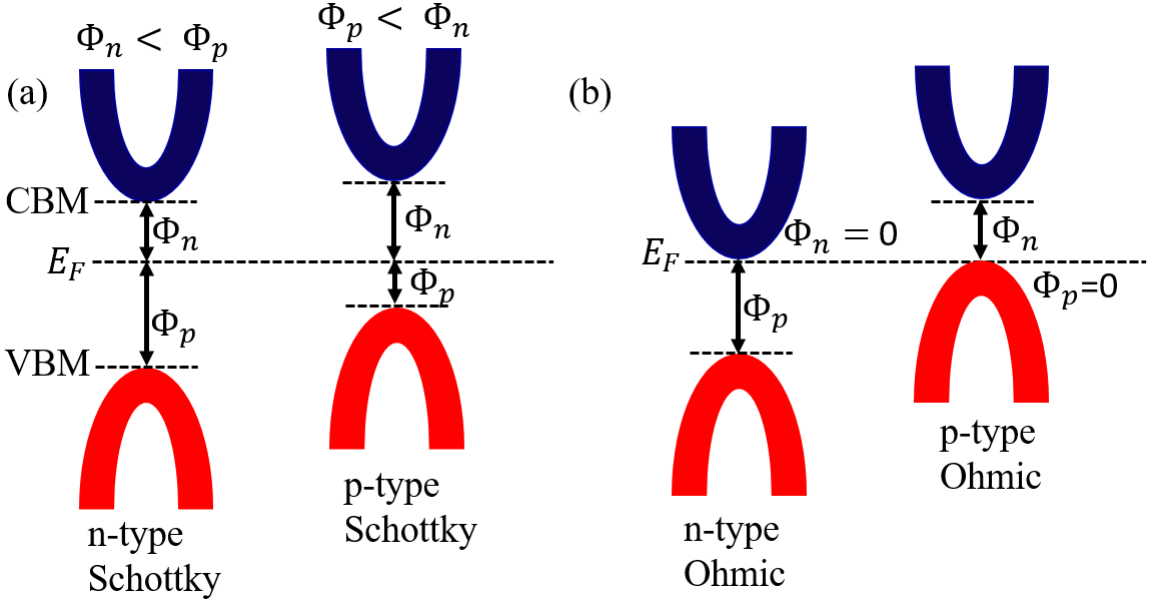}\caption{Schematic illustration of electrical contacts (Schottky and Ohmic)
formed at a metal-semiconductor interface.}
\label{fig:sbh} 
\end{figure}

The most interesting aspect of the C$_{6}$N$_{6}$/GE and hg-C$_{3}$N$_{4}$/GE
HTSs is that they form a metal-semiconductor contact, which can either
exhibit Schottky or Ohmic characteristics. To illustrate the distinction
between Schottky and Ohmic contacts more clearly, Fig.~\ref{fig:sbh}
shows schematic representations of these electrical contacts at a
metal-semiconductor interface. From the band-edge positions (VBM/CBM)
with respect to the Fermi levels as illustrated in Figs.~\ref{fig:EBS}(a)-\ref{fig:EBS}(b)
(GGA) and Figs.~\ref{fig:EBS}(d)-\ref{fig:EBS}(e) (HSE06), it is
easy to deduce whether the given HTS is Schottky-type or Ohmic in
nature. By comparing the band-edge positions of C$_{6}$N$_{6}$ and
hg-C$_{3}$N$_{4}$ layers with the respective Fermi levels of their
HTSs, we note that both C$_{6}$N$_{6}$/GE and hg-C$_{3}$N$_{4}$/GE
interfaces form Schottky contacts (SCs).

Further, we calculated the Schottky barrier heights (SBHs) for both C$_{6}$N$_{6}$/GE and hg-C$_{3}$N$_{4}$/GE HTSs. According to the Schottky-Mott rule~\cite{PhysRev.71.717},
the $n$-type SBH ($\Phi_{\mathrm{n}}$) and $p$-type SBH ($\Phi_{\mathrm{p}}$)
are defined as follows: 
\begin{align}
\Phi_{\mathrm{n}} & =\mathrm{E}_{\mathrm{CBM}}-\mathrm{E}_{\mathrm{F}},\label{eq:phin}\\
\Phi_{\mathrm{p}} & =\mathrm{E}_{\mathrm{F}}-\mathrm{E}_{\mathrm{VBM}}.\label{eq:phip}
\end{align}

\begin{figure*}[ht]
\includegraphics[width=1\linewidth]{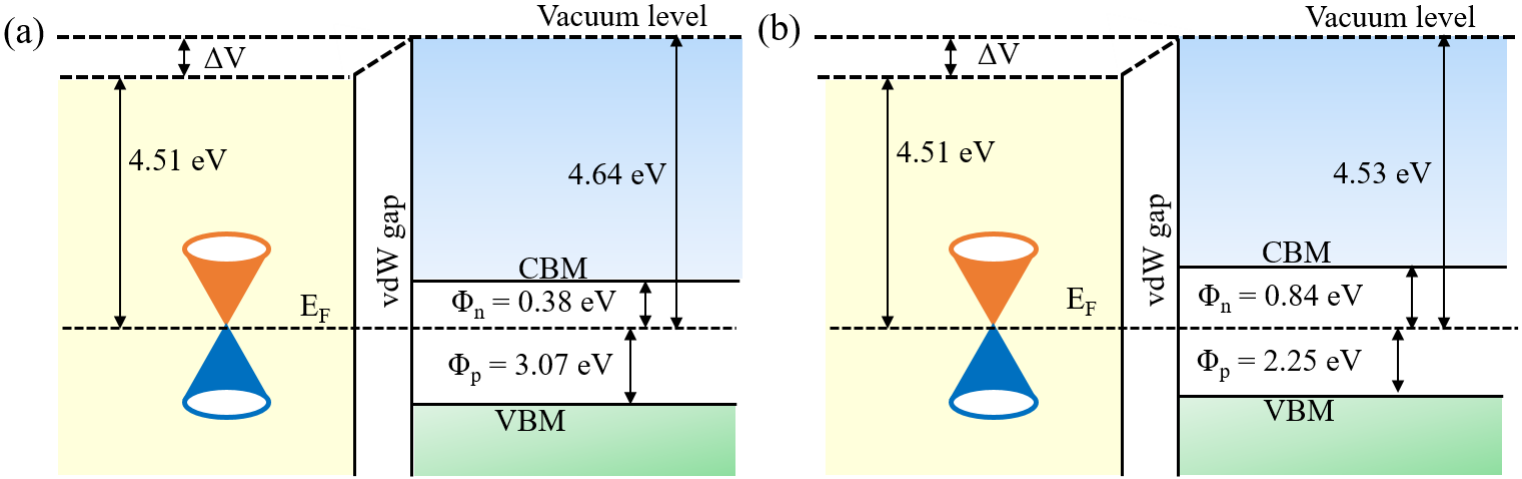}\caption{Schematic illustration of band alignment corresponding to the (a)
C$_{6}$N$_{6}$/GE and (b) hg-C$_{3}$N$_{4}$/GE heterostructures,
based on the HSE06 calculations. Both HTSs exhibit $n$-type Schottky contact behavior.}
\label{fig:bandbending} 
\end{figure*}

Above, $\Phi_{\mathrm{n}}$ and $\Phi_{\mathrm{p}}$ represent the
barrier heights for electrons and holes, respectively. E$_{\mathrm{CBM}}$
and E$_{\mathrm{VBM}}$ denote the CBM and VBM of the semiconductor
in the vdW HTSs. When the SBH is positive,
a Schottky contact forms, whereas if it is zero or negative, an Ohmic
contact forms. Additionally, the electronic band gap of the C$_{6}$N$_{6}$
and hg-C$_{3}$N$_{4}$ monolayers in the C$_{6}$N$_{6}$/GE and
hg-C$_{3}$N$_{4}$/GE HTSs, respectively, were calculated using $\mathrm{E}_{\mathrm{g}}=\mathrm{E}_{\mathrm{CBM}}-\mathrm{E}_{\mathrm{VBM}}$,
i.e., $\mathrm{E}_{\mathrm{g}}=\Phi_{\mathrm{n}}-\Phi_{\mathrm{p}}$.
In the band structure plots of both HTSs, the Fermi level is set to
zero, and the Dirac point of GE aligns with the Fermi level. Using
the Schottky-Mott rule, the barrier heights for the C$_{6}$N$_{6}$/GE
HTS are found to be $\Phi_{\mathrm{n}}$ = 0.38 eV and $\Phi_{\mathrm{p}}$
= 3.07 eV based on the HSE06 functional, while the PBE functional
yields values of 0.08 eV and 1.75 eV, respectively. For the hg-C$_{3}$N$_{4}$/GE
HTS, the barrier heights obtained using the HSE06 functional are $\Phi_{\mathrm{n}}$
= 0.84 eV and $\Phi_{\mathrm{p}}$ = 2.25 eV. The corresponding values
obtained from the PBE functional are 0.42 eV and 1.12 eV, respectively.
Consequently, $\Phi_{\mathrm{n}}<\Phi_{\mathrm{p}}$ for both HTSs using both PBE and HSE06 functionals,
indicating that they form $n$-SCs at the equilibrium
interlayer distance. The obtained SBHs and contact types for both HTSs, computed using PBE and HSE06 methods, are summarized in Table~\ref{tab:SBheight}. Moreover, the schematic
representation of band alignment for both the C$_{6}$N$_{6}$/GE
and hg-C$_{3}$N$_{4}$/GE HTSs corresponding to the results obtained
using the HSE06 method is shown in Fig.~\ref{fig:bandbending}.
\begin{table}[ht]
\centering
\caption{Schottky barrier heights of the C$_{6}$N$_{6}$/GE and hg-C$_{3}$N$_{4}$/GE heterostructures obtained from the Schottky-Mott rule, computed using the PBE and HSE06 methods, indicating $n$-type Schottky contacts ($n$-SCs) in both HTSs.}
\label{tab:SBheight}
\begin{ruledtabular}
\begin{tabular}{cccc}
HTS & $\Phi_{\mathrm{n}}$ (eV)   & $\Phi_{\mathrm{p}}$ (eV)   & Contact type \tabularnewline
 &  PBE/HSE06 & PBE/HSE06  &  \tabularnewline
\hline
C$_{6}$N$_{6}$/GE & 0.08 / 0.38 & 1.75 / 3.07 & $n$-SC \tabularnewline
hg-C$_{3}$N$_{4}$/GE & 0.42 / 0.84 & 1.12 / 2.25 & $n$-SC \tabularnewline
\end{tabular}
\end{ruledtabular}
\end{table}

\begin{figure}[ht]
\includegraphics[width=1\linewidth]{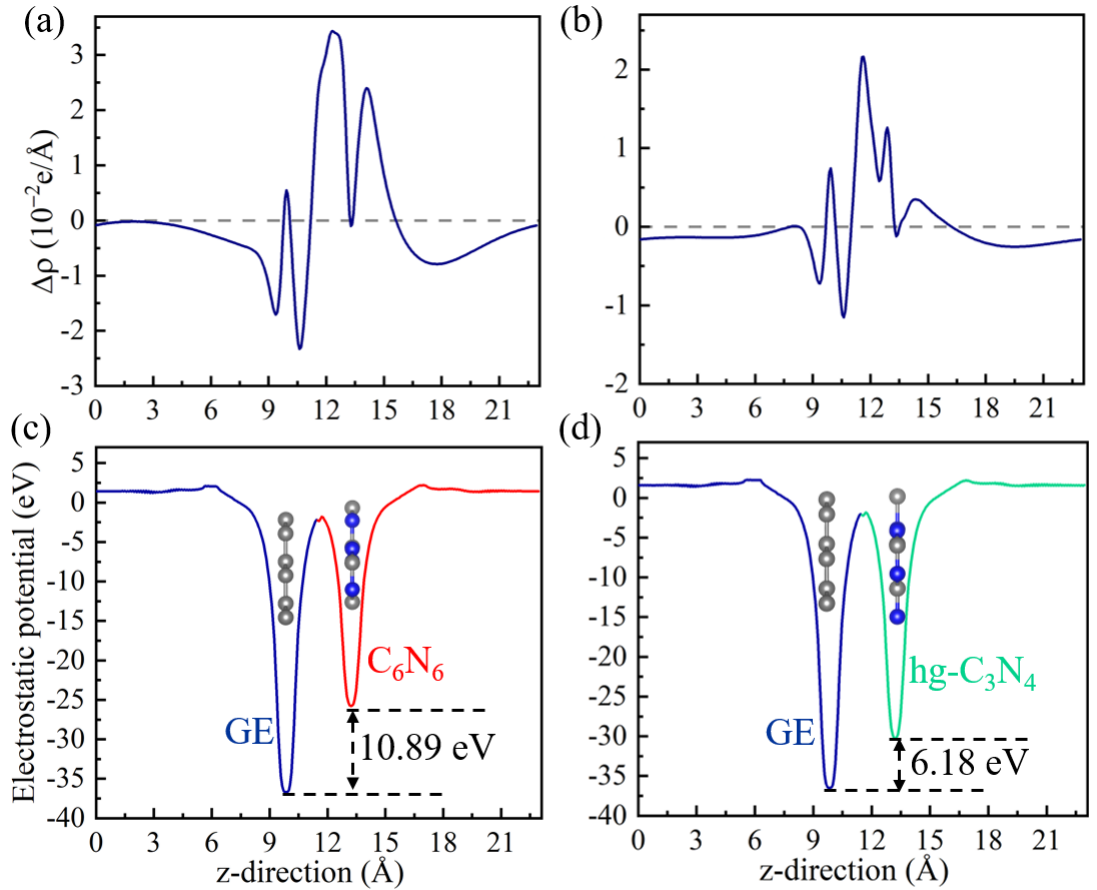}\caption{(a), (b) Calculated average electronic density difference ($\Delta\rho$)
and (c), (d) electrostatic potential for the C$_{6}$N$_{6}$/GE and
hg-C$_{3}$N$_{4}$/GE heterostructures, respectively, using the HSE06
functional.}
\label{fig:effec} 
\end{figure}

In order to understand the flow of charge transfer in the considered
HTSs, we further plotted the average electron density difference ($\Delta{\rho}$)
for C$_{6}$N$_{6}$/GE and hg-C$_{3}$N$_{4}$/GE HTSs using the
HSE06 functional, as depicted in Figs.~\ref{fig:effec}(a) and~\ref{fig:effec}(b),
respectively. The average electronic density difference for both the
HTSs reveals that the GE layer exhibits a negative value across the
interface of the HTS, indicating electron depletion, while the C$_{6}$N$_{6}$
and hg-C$_{3}$N$_{4}$ layers display positive values across the
interface of HTS, signifying electron accumulation. This indicates
that electrons transfer from the GE layer to the C$_{6}$N$_{6}$
and hg-C$_{3}$N$_{4}$ layers in both HTSs.

This charge redistribution is further supported by the electrostatic
potential profiles (Figs.~\ref{fig:effec}(c) and \ref{fig:effec}(d)),
which show that GE possesses a lower potential compared to the adjacent
semiconducting layers. The resulting potential gradient establishes
a built-in electric field directed from the 2D semiconductor toward
the GE layer. The resulting internal field polarizes electron wavefunctions
at the interface, giving rise to interfacial electric dipoles. The
presence of electric dipoles not only modulates the local band alignment
but also effectively reduces the charge injection barrier at the graphene
contact, thereby facilitating more efficient interfacial charge transfer.
Such interfacial characteristics are advantageous for enhancing the
performance of electronic and optoelectronic devices based on these
HTSs. Furthermore, to confirm the direction of charge transfer in
the C$_{6}$N$_{6}$/GE and hg-C$_{3}$N$_{4}$/GE HTSs, we calculated
the work functions of the individual monolayers. Using GGA, the calculated
work functions of GE, C$_{6}$N$_{6}$, and hg-C$_{3}$N$_{4}$ are
4.26 eV, 5.78 eV, and 4.63 eV, respectively. Using the HSE06 functional,
the corresponding values are 4.51 eV (GE), 7.49 eV (C$_{6}$N$_{6}$),
and 6.56 eV (hg-C$_{3}$N$_{4}$) (see Table~\ref{tab:parameter}).
It is evident that the GE monolayer consistently exhibits the lowest
work function among the three considered monolayers in both GGA and
HSE06 calculations. Moreover, this significant difference in work
function indicates that electrons will transfer from the GE layer
to the C$_{6}$N$_{6}$ and hg-C$_{3}$N$_{4}$ layers with relative
ease in their respective HTSs. These results further corroborate the
charge-redistribution behavior revealed by our calculations, and indicate that both C$_{6}$N$_{6}$/GE and hg-C$_{3}$N$_{4}$/GE HTSs are promising candidates for high-efficiency field-effect transistor (FET)
applications.

\subsubsection{Modulation of Schottky barrier height and contact type under perpendicular field and vertical strain}
The ability to tune the SBH and contact type is crucial for the realization of advanced Schottky devices. In this context, we explore the tunability of SBH and contact types in the C$_{6}$N$_{6}$/GE and hg-C$_{3}$N$_{4}$/GE
HTSs. Previous studies have demonstrated that the SBH and contact
types in GE-based HTSs can be significantly altered by varying the
interlayer distance or by applying an external perpendicular electric
field ($\mathrm{E}_{\perp}$)~\cite{PhysRevLett.114.066803,nguyen2020interlayer}.
\begin{figure}[ht]
\includegraphics[width=1\linewidth]{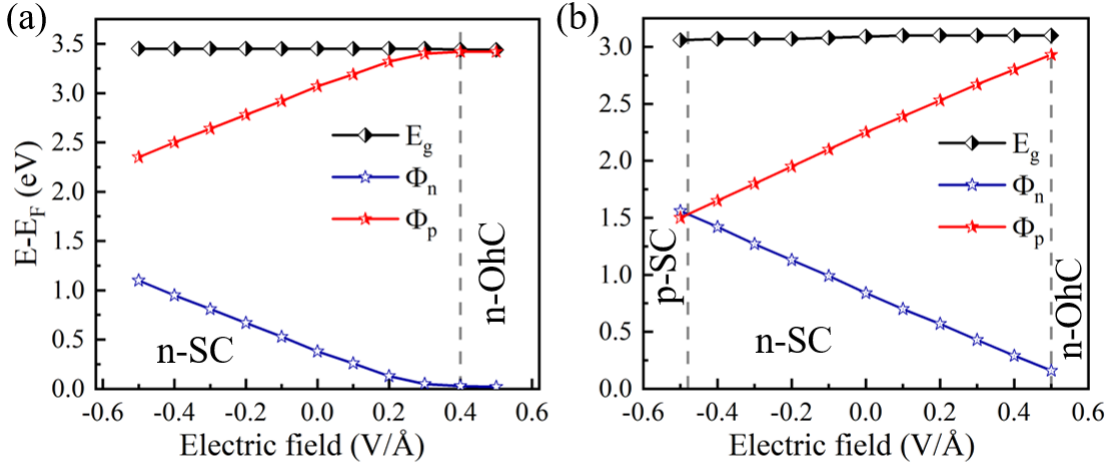}\caption{Variation of the Schottky barrier height (SBH) under an external static perpendicular electric
field ($\mathrm{E}_{\perp}$) for (a) C$_{6}$N$_{6}$/GE and (b) hg-C$_{3}$N$_{4}$/GE
heterostructures, calculated using the HSE06 functional. The gray
dotted line denotes the transition in contact type in the respective
heterostructure.}
\label{fig:Efield} 
\end{figure}

Therefore, we first examine the impact of the $\mathrm{E}_{\perp}$
on the electronic properties and contact types of the GE-based HTSs,
followed by a discussion of the effects of varying the interlayer
distance. We applied a $\mathrm{E}_{\perp}$ along the $z$-axis,
where the field direction is defined as positive when oriented from
the GE monolayer toward the C$_{n}$N$_{m}$ monolayer, with the negative
direction being just opposite to it. The influence of $\mathrm{E}_{\perp}$
on the SBHs ($\Phi_{\mathrm{n}}$ and $\Phi_{\mathrm{p}}$) and the
band gap ($\mathrm{E}_{\mathrm{g}}$) for the C$_{6}$N$_{6}$/GE
and hg-C$_{3}$N$_{4}$/GE HTSs is depicted in Figs.~\ref{fig:Efield}(a)-\ref{fig:Efield}(b),
respectively, obtained from HSE06 calculations.

In case of C$_{6}$N$_{6}$/GE HTS (Fig.~\ref{fig:Efield}(a)), with
increasing $\mathrm{E}_{\perp}$, $\Phi_{\mathrm{p}}$ increases steadily,
while $\Phi_{\mathrm{n}}$ decreases, as a result of which $\Phi_{\mathrm{n}}<\Phi_{\mathrm{p}}$,
indicating that the HTS retains its $n$-SC character throughout the
range of applied $\mathrm{E}_{\perp}$. However, when $\mathrm{E}_{\perp}$
surpasses approximately +0.4 V/\AA, $\Phi_{\mathrm{n}}$ approaches
zero, indicating a transition from an $n$-SC to an $n$-type Ohmic contact ($n$-OhC). We can understand these results based on the projected band structures presented in Figs. S3(a)-S3(e)
of the SM~\cite{SI}, from which it is obvious that, with the increasing $\mathrm{E}_{\perp}$, the Fermi level shifts toward the CBM of the HTS, leading to a decrease in $\Phi_{\mathrm{n}}$ and an increase in $\Phi_{\mathrm{p}}$, ultimately
resulting in a transition from an $n$-SC to an $n$-OhC when the field
reaches the critical value of $+0.4$~V/\AA. Further, from our GGA
calculations, for C$_{6}$N$_{6}$/GE HTS, a similar trend is obtained
(see Fig. S4(a) of the SM~\cite{SI}).

For the hg-C$_{3}$N$_{4}$/GE HTS, as shown in Fig.~\ref{fig:Efield}(b), the application of an increasing E$_{\perp}$, in this HTS as well, leads to an increase in $\Phi_{\mathrm{p}}$, and a decrease in $\Phi_{\mathrm{n}}$. For
E$_{\perp}\leq-0.48$V/\AA, $\Phi_{\mathrm{p}}<\Phi_{\mathrm{n}}$,
therefore, in that region of E$_{\perp}$, the HTS behaves as a $p$-type
Schottky contact ($p$-SC). But, when E$_{\perp}$$>-0.48$V/\AA, $\Phi_{\mathrm{n}}<\Phi_{\mathrm{p}}$,
leading to a transition to an $n$-SC. Furthermore, when the positive
E$_{\perp}$ exceeds +0.5 V/\AA, $\Phi_{\mathrm{n}}$ tends toward
zero, signifying a transition from $n$-SC to $n$-OhC. These trends are
understandable on examining the E$_{\perp}$ dependent projected band
structures presented in Figs. S3(f)-S3(j)
of the SM~\cite{SI}: (a) a negative E$_{\perp}$ brings the VBM closer to the
Fermi level, while the CBM shifts further away, leading to a reduction
in $\Phi_{\mathrm{p}}$ and an increase in $\Phi_{\mathrm{n}}$. At
E$_{\perp}$ $\leq$ -0.48 V/\AA, the VBM becomes so close to the
Fermi level that $\Phi_{\mathrm{p}}$ becomes smaller than $\Phi_{\mathrm{n}}$,
leading to a transition from an $n$-SC to a $p$-SC, and, (b) in contrast,
under a positive E$_{\perp}$, the CBM approaches the Fermi level,
while the VBM moves farther away from it, resulting in an increase
of $\Phi_{\mathrm{p}}$ and a decrease in $\Phi_{\mathrm{n}}$. At
a critical field of $+0.5$~V/\AA, this behavior induces a transition
from an $n$-SC to an $n$-OhC, similar to the case of C$_{6}$N$_{6}$/GE
HTS, demonstrating the field-tunable electronic properties of the
hg-C$_{3}$N$_{4}$/GE HTS.

From our GGA calculations (see Fig. S4(b) of the SM~\cite{SI}),
we notice the same overall field-dependent trends as in the HSE06
results, with one notable quantitative difference: the $p$-SC to $n$-SC transition occurs at a smaller magnitude of the applied negative electric field. Specifically, the hg-C$_{3}$N$_{4}$/GE HTS behaves as a $p$-SC for E$_{\perp}\leq-0.26$ V/\AA, consistent
with the reduction of $\Phi_{\mathrm{p}}$ and increase in $\Phi_{\mathrm{n}}$
under negative E$_{\perp}$. For E$_{\perp}>-0.26$ V/\AA, $\Phi_{\mathrm{n}}$
becomes smaller than $\Phi_{\mathrm{p}}$, indicating a transition
to an $n$-SC. Apart from this shift in the critical field value,
the qualitative evolution of the Schottky barriers with E$_{\perp}$
remains unchanged, confirming that both GGA and HSE06 predict the
same field-tunable behavior of the hg-C$_{3}$N$_{4}$/GE HTS.

These findings highlight the effective tunability of the SBHs and
contact types in the C$_{6}$N$_{6}$/GE and hg-C$_{3}$N$_{4}$/GE
HTSs through the application of an $\mathrm{E}_{\perp}$. Further,
the electronic band gap ($\mathrm{E}_{\mathrm{g}}$) of both the HTSs
remains nearly constant throughout the applied $\mathrm{E}_{\perp}$,
demonstrating the robust electronic properties of both HTSs.

\begin{figure}[ht]
\includegraphics[width=1\linewidth]{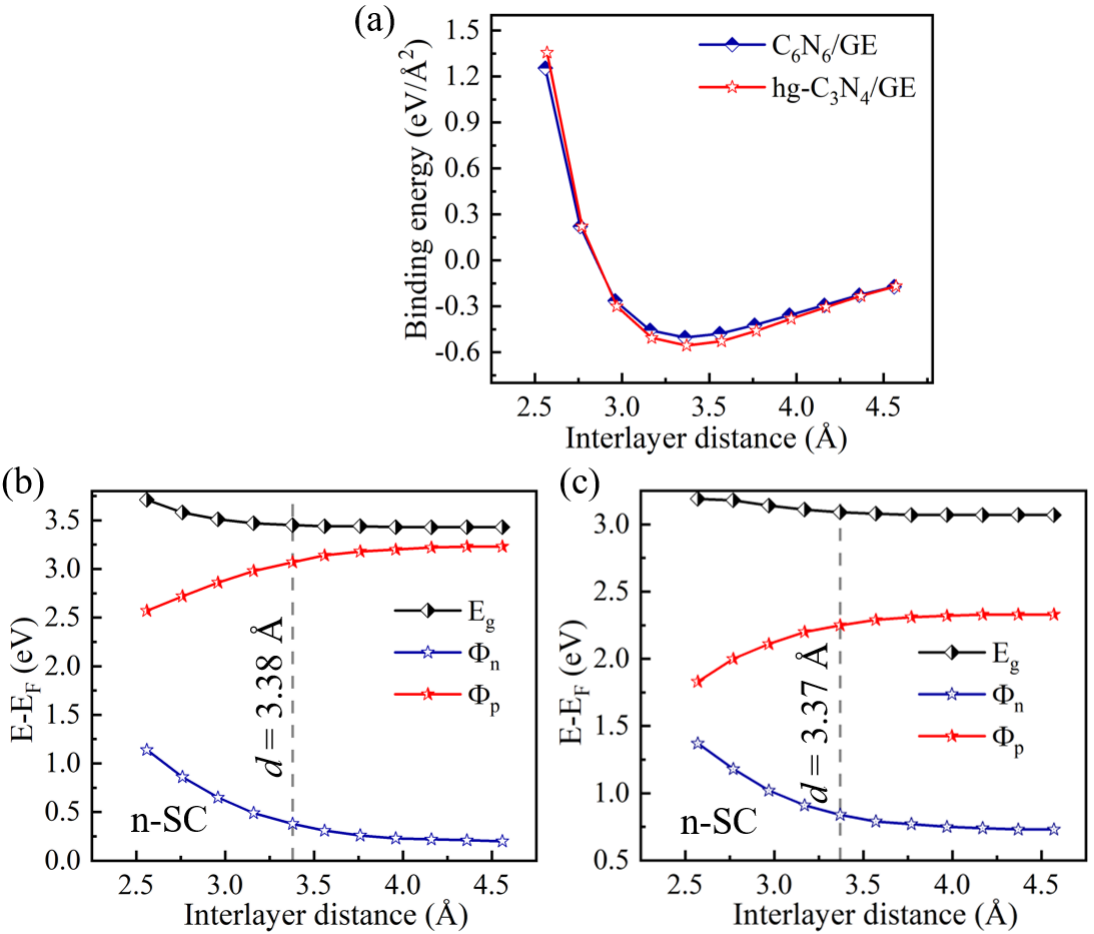}\caption{(a) Variation of binding energy with interlayer distance for C$_{6}$N$_{6}$/GE
and hg-C$_{3}$N$_{4}$/GE heterostructures, obtained using GGA. (b),
(c) Variation of Schottky barrier height (SBH) with interlayer distance
for C$_{6}$N$_{6}$/GE and hg-C$_{3}$N$_{4}$/GE HTSs, respectively,
were obtained using the HSE06 functional. The gray dotted line denotes
the equilibrium distance obtained in the respective heterostructure.}
\label{fig:inter} 
\end{figure}
Next, we study the variation of SBH and contact type in C$_{6}$N$_{6}$/GE
and hg-C$_{3}$N$_{4}$/GE HTSs as a function of interlayer distance. Fig.~\ref{fig:inter}(a) shows the variation in binding energy with interlayer distance for both C$_{6}$N$_{6}$/GE and hg-C$_{3}$N$_{4}$/GE HTSs, calculated using GGA. It is evident that the binding energy reaches its minimum at the equilibrium interlayer distances of 3.38~\AA~for C$_{6}$N$_{6}$/GE and 3.37~\AA~for hg-C$_{3}$N$_{4}$/GE (see Table~\ref{tab:parofhts}), indicating optimal stability at these separations. Further, as shown in Figs.~\ref{fig:inter}(b) and \ref{fig:inter}(c), the SBH and contact type, obtained from HSE06 calculations, can be modulated by altering the interlayer coupling.

For the C$_{6}$N$_{6}$/GE HTS (Fig.~\ref{fig:inter}(b)), when the interlayer distance ($d$) decreases below the equilibrium value, $\Phi_{\mathrm{n}}$ increases while $\Phi_{\mathrm{p}}$ decreases. Nevertheless, $\Phi_{\mathrm{n}}$ remains smaller than $\Phi_{\mathrm{p}}$ down to $d = 2.56$~\AA, indicating the formation of an $n$-SC. Conversely, for the interlayer distance larger than the equilibrium separation, $\Phi_{\mathrm{p}}$ increases and $\Phi_{\mathrm{n}}$ decreases, but still $\Phi_{\mathrm{n}}$ remains lower than $\Phi_{\mathrm{p}}$ ($\Phi_{\mathrm{n}} < \Phi_{\mathrm{p}}$), indicating the persistence of the $n$-SC character.
This can further be understood from the projected $d$-dependent band structure of C$_{6}$N$_{6}$/GE HTS (see Figs. S5(a)-S5(e) of the SM~\cite{SI}), in that, with decreasing $d$ compared to the equilibrium $d$ of
the HTS, the Fermi level moves closer to the VBM, leading to an increase in $\Phi_{\mathrm{n}}$, and a decrease in $\Phi_{\mathrm{p}}$. In contrast, with increasing $d$ beyond the equilibrium value of the HTS, the Fermi level moves away from the VBM, leading to an increase in $\Phi_{\mathrm{p}}$, and a decrease in $\Phi_{\mathrm{n}}$. However, throughout the range of $d$,  $\Phi_{\mathrm{n}}$ is noticed to be lower than $\Phi_{\mathrm{p}}$, therefore the HTS consistently exhibits an $n$-SC character. 

Within the GGA framework, a similar trend is noticed for interlayer distances smaller than the equilibrium separation (see Fig. S4(c) of the SM~\cite{SI}), where the HTS exhibits $n$-SC behavior. However, for the interlayer distances larger than the equilibrium separation, a different behavior emerges. In particular, at an interlayer distance of 3.56~\AA, the SBH, $\Phi_{\mathrm{n}}$, approaches zero (0.05 eV), indicating the formation of an $n$-Ohc. This is in contrast to the HSE06 results, where the HTS continues to exhibit $n$-SC  character, as $\Phi_{\mathrm{n}}$ remains finite (0.19 eV at 4.56~\AA) in the HSE06 calculations.
 
In the case of hg-C$_{3}$N$_{4}$/GE HTS (see Fig.~\ref{fig:inter}(c)), the SBH exhibits a similar behavior across all interlayer distances,
calculated using the HSE06 functional. When $d$ is reduced from the
equilibrium distance, $\Phi_{\mathrm{p}}$ decreases and $\Phi_{\mathrm{n}}$
increases, but still $\Phi_{\mathrm{n}}$ remains lower than $\Phi_{\mathrm{p}}$
up to the studied interlayer distance of 2.57~\AA, and hence exhibits $n$-SC character. Moreover, for the distance greater than
the equilibrium distance, $\Phi_{\mathrm{n}}$ consistently remains
lower than $\Phi_{\mathrm{p}}$, maintaining again the $n$-SC in
the hg-C$_{3}$N$_{4}$/GE HTS. From the projected band structures
(see Figs. S5(f)-S5(j) of the SM~\cite{SI}), we note that a reduction in the interlayer distance below the equilibrium value causes the Fermi level to approach the VBM, resulting in an increase in $\Phi_{\mathrm{n}}$ and a decrease in $\Phi_{\mathrm{p}}$. On the other hand, on increasing $d$ beyond the equilibrium value shifts the Fermi level closer to the CBM of the HTS, leading to an increase in $\Phi_{\mathrm{p}}$ and a decrease in $\Phi_{\mathrm{n}}$, so that for all these values of $d$, $\Phi_{\mathrm{n}}<\Phi_{\mathrm{p}}$,
maintaining its $n$-SC character.
Within the GGA framework (see Fig. S4(d) of SM~\cite{SI}), when the interlayer distance is less than 2.7 $\text{\AA}$, $\Phi_{\mathrm{p}}<\Phi_{\mathrm{n}}$, indicating a $p$-SC. However, for distances greater than 2.7~\AA, $\Phi_{\mathrm{n}}<\Phi_{\mathrm{p}}$, due to this the HTS will have an $n$-SC. 

Further, the band gap exhibits a slight variation with changes in the interlayer distance for both the HTSs (see Fig.~\ref{fig:inter}(b)-(c); Figs. S4(c)-S4(d) of the SM~\cite{SI}). As the interlayer distance decreases from the equilibrium distance (3.38~\AA~for C$_{6}$N$_{6}$/GE and 3.37~\AA~for hg-C$_{3}$N$_{4}$/GE), the band gap gradually increases. Conversely, as the interlayer distance increases beyond the equilibrium value, the band gap stabilizes at approximately 3.43 eV (1.9 eV) for C$_{6}$N$_{6}$/GE and 2.33 eV (1.8 eV) for hg-C$_{3}$N$_{4}$/GE within the HSE06 (GGA) framework.

\subsubsection{Impact of electric field and strain on the electronic properties
of C$_{6}$N$_{6}$/hg-C$_{3}$N$_{4}$ heterostructure}

\begin{figure}[ht]
\includegraphics[width=1\linewidth]{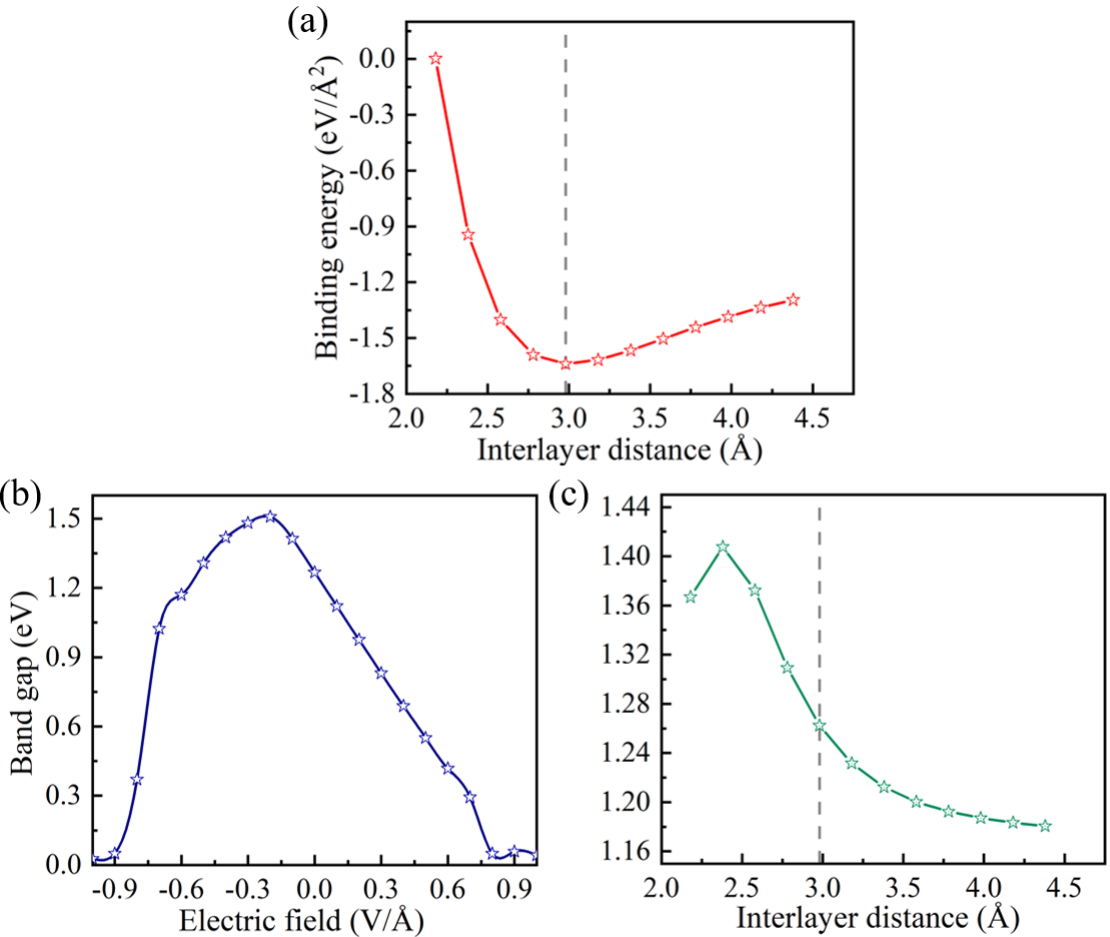}\caption{Variation of (a) binding energy with interlayer distance, (b) electronic
band gap with external perpendicular electric field ($\mathrm{E}_{\perp}$),
and (c) electronic band gap with interlayer distance for the C$_{6}$N$_{6}$/hg-C$_{3}$N$_{4}$
heterostructure, calculated using GGA.}
\label{fig:varai} 
\end{figure}

Next, we investigate the electronic properties of the C$_{6}$N$_{6}$/hg-C$_{3}$N$4$
HTS, composed of two semiconducting monolayers, with respect to variations
in the $\mathrm{E}_{\perp}$, and the interlayer distance. In our calculations, the equilibrium
interlayer distance of the C$_{6}$N$_{6}$/hg-C$_{3}$N$_{4}$ HTS
is found to be 2.98~{\AA} (see Table~\ref{tab:parofhts}), which
is consistent with an earlier first-principles study on C$_{6}$N$_{6}$/hg-C$_{3}$N$_{4}$
HTS~\cite{liang2016photocatalytic}, where interlayer distances in
the range of 2.63-3.02~{\AA} were reported depending on the stacking
configuration. Fig.~\ref{fig:varai}(a) illustrates the variation
of binding energy with the interlayer distance, showing that the minimum
binding energy occurs at the equilibrium distance of 2.98 \AA, which
corresponds to the energetically most stable configuration of the
HTS.

Further, Figs.~\ref{fig:varai}(b) and~\ref{fig:varai}(c) illustrate
that the electronic band gap of the C$_{6}$N$_{6}$/hg-C$_{3}$N$_{4}$
HTS is strongly influenced by the application of an $\mathrm{E}_{\perp}$,
and the variation in the interlayer distance, respectively. As $\mathrm{E}_{\perp}$
increases from -1.0 V/\AA~to +1.0 V/\AA~(see Fig.~\ref{fig:varai}(b)),
the band gap first increases, reaching a maximum of 1.51 eV at -0.2
V/\AA, and then gradually decreases, becoming nearly zero at both
the extremes of the $\mathrm{E}_{\perp}$ (0.03 eV at -1.0 V/\AA~and
0.04 eV at +1.0 V/\AA). The electronic band structure plots for the
selected $\mathrm{E}_{\perp}$ values are shown in Figs. S6(a)-S6(e)
(top panels) of the SM~\cite{SI}. It is noticed that at the equilibrium configuration
($d$ = 0~{\AA} and $\mathrm{E}_{\perp}$ = 0), the VBM is predominantly
derived from the hg-C$_{3}$N$_{4}$ layer, whereas the CBM primarily
originates from the C$_{6}$N$_{6}$ layer. This indicates that the
C$_{6}$N$_{6}$/hg-C$_{3}$N$_{4}$ vdW HTS exhibits a type-II band
alignment, which is favorable for efficient electron-hole separation~\cite{liang2016photocatalytic}.
At the negative value $\mathrm{E}_{\perp}$ (-0.2 V/\AA), the VBM
of the C$_{6}$N$_{6}$ layer shifts toward the Fermi level, while
that of hg-C$_{3}$N$_{4}$ shifts away (see Fig. S6
of the SM~\cite{SI}), inducing a transition from type-II to type-I band alignment.
The HTS continues to exhibit type-I alignment at $\mathrm{E}_{\perp}$
= -0.3 V/\AA. Interestingly, as the magnitude of the negative field is further increased
(-0.4 V/\AA) , the CBM of hg-C$_{3}$N$_{4}$ continuously shifts
downward, while that of C$_{6}$N$_{6}$ shifts upward, becoming energetically
higher. In this case, the VBM and CBM are dominated by the C$_6$N$_6$ and hg-C$_3$N$_4$ layers, respectively, leading to a transition back to type-II alignment. For all the applied positive $\mathrm{E}_{\perp}$, the HTS continues to exhibit robust type-II alignment. It should be noted that the electric field range considered here is experimentally achievable via dual ionic gating in two-dimensional materials and heterostructures~\cite{domaretskiy2022quenching,weintrub2022generating}.

Further, Fig.~\ref{fig:varai}(c) shows the band gap variation of
C$_{6}$N$_{6}$/hg-C$_{3}$N$_{4}$ HTS with the interlayer distance.
The band gap is noticed to be larger than that at the equilibrium
position when the interlayer distance is reduced below the equilibrium
value. On the other hand, increasing the interlayer distance beyond
the equilibrium value leads to a decrease in the band gap compared
to that at equilibrium. Both these results are consistent with the
quantum confinement effects. The corresponding band structure plots
for selected interlayer distances are shown in Figs. S6(f)-S6(j) of the SM~\cite{SI}. Notably, at $d$ = 2.18 \AA, both
the VBM and CBM of C$_{6}$N$_{6}$/hg-C$_{3}$N$_{4}$ HTS originate
from the C$_{6}$N$_{6}$ layer, thus indicating a transition from
type-II to type-I band alignment. These results collectively highlight
the controllable transitional behavior between type-II and type-I
band alignment in the C$_{6}$N$_{6}$/hg-C$_{3}$N$_{4}$ HTS via
external stimuli such as $\mathrm{E}_{\perp}$ and interlayer distance,
making it a promising candidate for nanoscale electronic and optoelectronic
devices.

\subsection{Optical properties}
In the present work, the optical spectra are computed
using RPA, which includes screening
and local-field effects, without explicitly including the electron-hole
interactions. Additionally, BSE calculations were performed on top
of single-shot G$_{0}$W$_{0}$ quasiparticle corrections to evaluate the exciton binding energies of only the gapped systems, because strong dielectric screening in metallic systems prevents the formation of bound electron-hole pairs, rendering the concept of bound excitons meaningless for the metallic ones.
\begin{figure*}[ht]
\textcolor{blue}{\includegraphics[width=1\linewidth]{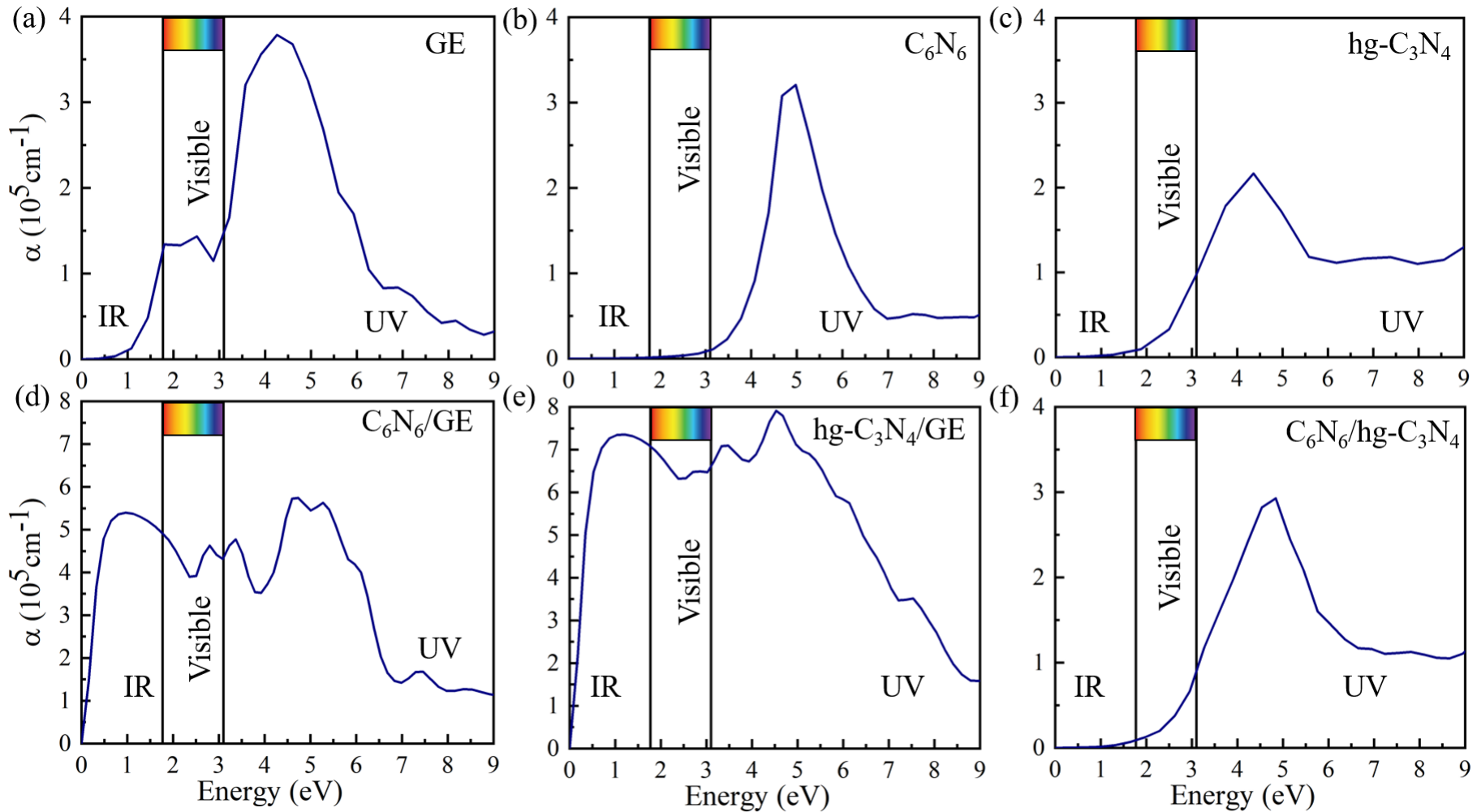}}\caption{Calculated absorption coefficient ($\alpha$, cm$^{-1}$) for the
isolated monolayers (a) GE, (b) C$_{6}$N$_{6}$, and (c) hg-C$_{3}$N$_{4}$,
and for the corresponding heterostructures (d) C$_{6}$N$_{6}$/GE,
(e) hg-C$_{3}$N$_{4}$/GE, and (f) hg-C$_{3}$N$_{4}$/C$_{6}$N$_{6}$,
computed using the RPA on top of GGA.}
\label{fig:opti} 
\end{figure*}
Fig.~\ref{fig:opti} shows the optical absorption spectra of the
isolated monolayers along with those of the considered HTSs. Distinct
spectral features are observed for each structure. We will discuss
these features for the monolayers, followed by the HTSs. The GE monolayer
(Fig.~\ref{fig:opti}(a)), consistent with its well-known optical
response, shows absorption both in the infrared (IR) and visible regions,
with a pronounced absorption peak at 4.25 eV in the ultraviolet (UV)
corresponding to the $\pi$-plasmon. These results of ours are in
excellent agreement with those reported by other authors~\cite{rafique2017manipulating,muhammad2017first,rani2014dft},
except for the high-energy $\sigma+\pi$ plasmonic peak, which we
did not target in this work. The C$_{6}$N$_{6}$ and hg-C$_{3}$N$_{4}$
monolayers exhibit prominent UV absorption, with the C$_{6}$N$_{6}$
monolayer showing a sharp plasmonic peak at 4.98 eV (Fig.~\ref{fig:opti}(b)),
while hg-C$_{3}$N$_{4}$ displays a dominant peak at 4.36 eV (Fig.~\ref{fig:opti}(c))
also corresponding to the $\pi$-plasmons. Our results on C$_{6}$N$_{6}$ monolayer are in excellent agreement with previous computational works both on the regions of negligible absorption as well as for the location
of the intense plasmonic peak~\cite{sun2020first,zhao2022promising,pang2024first}.

Moreover, previous theoretical and experimental investigations on
pristine hg-C$_{3}$N$_{4}$ monolayer consistently report that its
optical response is dominated by absorption in the UV region, while
the absorption in the visible range remains weak~\cite{anjum2025first,makaremi2018band,che2020plasmonic}.
This behavior has been widely discussed as a key limitation for visible-light
photocatalysis and forms the basis for band-engineering strategies
such as doping, or HTS formation~\cite{anjum2025first,makaremi2018band}.
Additionally, Che~\\textit{et al}.~\cite{che2020plasmonic} performed experimental
UV-visible measurements on pristine hg-C$_{3}$N$_{4}$ monolayer,
reporting absorption predominantly in the 250-450 nm range (2.75-4.96
eV), which arises from charge-transfer transitions between N-2$p$
valence states and C-2$p$ conduction states. Although the experimental
spectra exhibit an absorption edge extending into the visible region,
the strongest absorption remains in the UV regime. This observation
is in good agreement with our theoretical results, confirming that
the dominant optical response of the hg-C$_{3}$N$_{4}$ monolayer
is UV-driven, with comparatively weak visible-light absorption. The
consistency between these reported trends and our calculated absorption
spectrum further validates the reliability of our results for the
hg-C$_{3}$N$_{4}$ monolayer.

In the C$_{6}$N$_{6}$/GE HTS (Fig.~\ref{fig:opti}(d)), a pronounced
absorption feature is noticed at 0.97 eV in the IR region, while a
distinct peak appears at 2.80 eV in the visible region. Additional
absorption peaks at 3.37 eV, 4.74 eV, and 5.28 eV are found in the
UV region, highlighting the broad absorption capability of the HTS.
A similar trend is seen for the hg-C$_{3}$N$_{4}$/GE HTS, where
the first absorption peak arises at 1.22 eV in the IR, with two pronounced
peaks at 3.49 eV and 7.92 eV noticed in the UV region (Fig.~\ref{fig:opti}(e)).
The hg-C$_{3}$N$_{4}$/C$_{6}$N$_{6}$ HTS (Fig.~\ref{fig:opti}(f))
exhibits a dominant peak at 4.84 eV in the UV region, broadly in the
same region as in the monolayer C$_{6}$N$_{6}$, indicating its suitability
for the UV-based applications. Thus, the HTSs show enhanced absorption
across an extended energy range, making them promising candidates
for high-performance optical and optoelectronic devices.

Further, we investigated the refractive index ($n(\omega)$), extinction
coefficient (K$(\omega)$), electron energy-loss function (L$(\omega)$),
and reflectivity (R$(\omega)$) of the isolated monolayers as well
as for the considered HTSs.

\begin{figure*}[ht]
\includegraphics[width=1\linewidth]{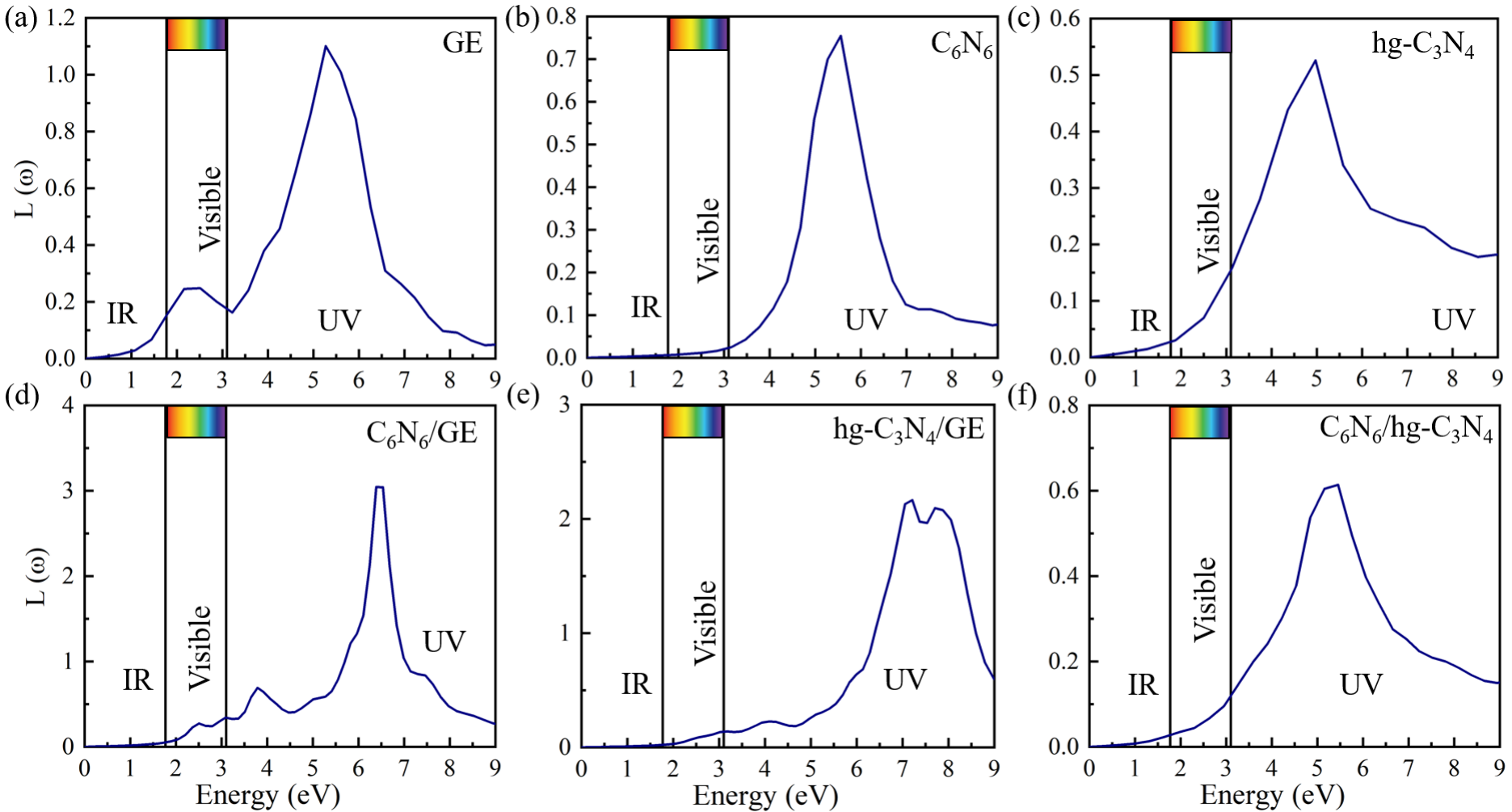}\caption{Calculated electron energy-loss function, $L(\omega)$, for the isolated
monolayers (a) GE, (b) C$_{6}$N$_{6}$, and (c) hg-C$_{3}$N$_{4}$,
and for the corresponding heterostructures (d) C$_{6}$N$_{6}$/GE,
(e) hg-C$_{3}$N$_{4}$/GE, and (f) hg-C$_{3}$N$_{4}$/C$_{6}$N$_{6}$,
calculated using the RPA approach.}
\label{fig:optiE} 
\end{figure*}

Fig.~\ref{fig:optiE} shows the calculated electron energy-loss function
(L$(\omega)$) for the isolated monolayers and the corresponding HTSs.
In addition to the interband excitations, L$(\omega)$ provides important
information about the collective electronic excitations, i.e., plasmons,
of the studied systems. The calculated $L(\omega)$ of the GE monolayer
(Fig.~\ref{fig:optiE}(a)) exhibits characteristic features within
the visible and UV energy regions. A small peak is obtained at 2.51~eV
in the visible region, while a pronounced peak at 5.27~eV dominates
the UV region of the spectrum and can be attributed to the $\pi$-plasmon
excitation. These prominent peaks are in good agreement with previously
reported experimental and theoretical results~\cite{li2023first,rost2023efficient}.
In addition, for the C$_{6}$N$_{6}$ (Fig.~\ref{fig:optiE}(b))
and hg-C$_{3}$N$_{4}$ (Fig.~\ref{fig:optiE}(c)) monolayers, we
obtained prominent peaks in the UV spectrum at 5.56 eV and 4.97 eV,
respectively, corresponding to their $\pi$-plasmons. For the C$_{6}$N$_{6}$/GE
HTS (Fig.~\ref{fig:optiE}(d)), we obtained a peak at 2.51 eV in
the visible spectrum, accompanied by additional UV features at 3.79~eV
and 6.54~eV. The hg-C$_{3}$N$_{4}$/GE HTS (Fig.~\ref{fig:optiE}(e))
exhibits two pronounced UV peaks at 7.21~eV and 7.71~eV, whereas
the hg-C$_{3}$N$_{4}$/C$_{6}$N$_{6}$ HTS (Fig.~\ref{fig:optiE}(f))
shows a dominant peak at 5.45~eV. These changes indicate a strong
influence of interlayer coupling on the plasmonic response of the
HTS.

Furthermore, we computed additional optical properties, namely, the
extinction coefficient (K$(\omega)$), the refractive index (n$(\omega)$),
and the reflectivity (R$(\omega)$), with the results presented in
Figs. S7-S9 of the SM~\cite{SI}. Fig. S7 of the SM~\cite{SI}
shows the extinction coefficient ($K(\omega)$) for the monolayers
and their corresponding HTS. For the GE monolayer (see Fig. S7(a) of the SM~\cite{SI}),
$K(\omega)$ exhibits a strong low-energy response with pronounced
peaks in the IR region. With increasing photon energy, the extinction
coefficient decreases through the visible range, followed by a broad
enhancement in the UV region around 3.57-4.26~eV, after which it
gradually diminishes. For the C$_{6}$N$_{6}$ monolayer (see Fig. S7(b) of the SM~\cite{SI}),
$K(\omega)$ remains negligible in the IR and visible regions, while
a distinct and dominant peak appears in the UV region at 4.68~eV.
Similarly, the hg-C$_{3}$N$_{4}$ monolayer (see Fig. S7(c) of the SM~\cite{SI})
shows weak optical attenuation at low photon energies, followed by
a gradual increase across the visible region and a broad maximum in
the UV range around 3.74-4.36~eV. In the HTSs, a pronounced enhancement
of optical attenuation is noticed at low photon energies. The C$_{6}$N$_{6}$/GE
HTS (see Fig. S7(d) of the SM~\cite{SI}) exhibits very large values of $K(\omega)$
in the IR region, which decreases rapidly with increasing energy,
leading to suppressed attenuation in the visible and UV regions. A
similar trend is noticed for the hg-C$_{3}$N$_{4}$/GE HTS (see Fig. S7(e) of the SM~\cite{SI}).
In contrast, the C$_{6}$N$_{6}$/hg-C$_{3}$N$_{4}$ HTS (see Fig. S7(f) of the SM~\cite{SI})
displays a moderate low-energy response, followed by a pronounced
peak in the UV region around 4.53-4.84~eV.
\begin{figure*}[ht]
\includegraphics[width=0.58\linewidth]{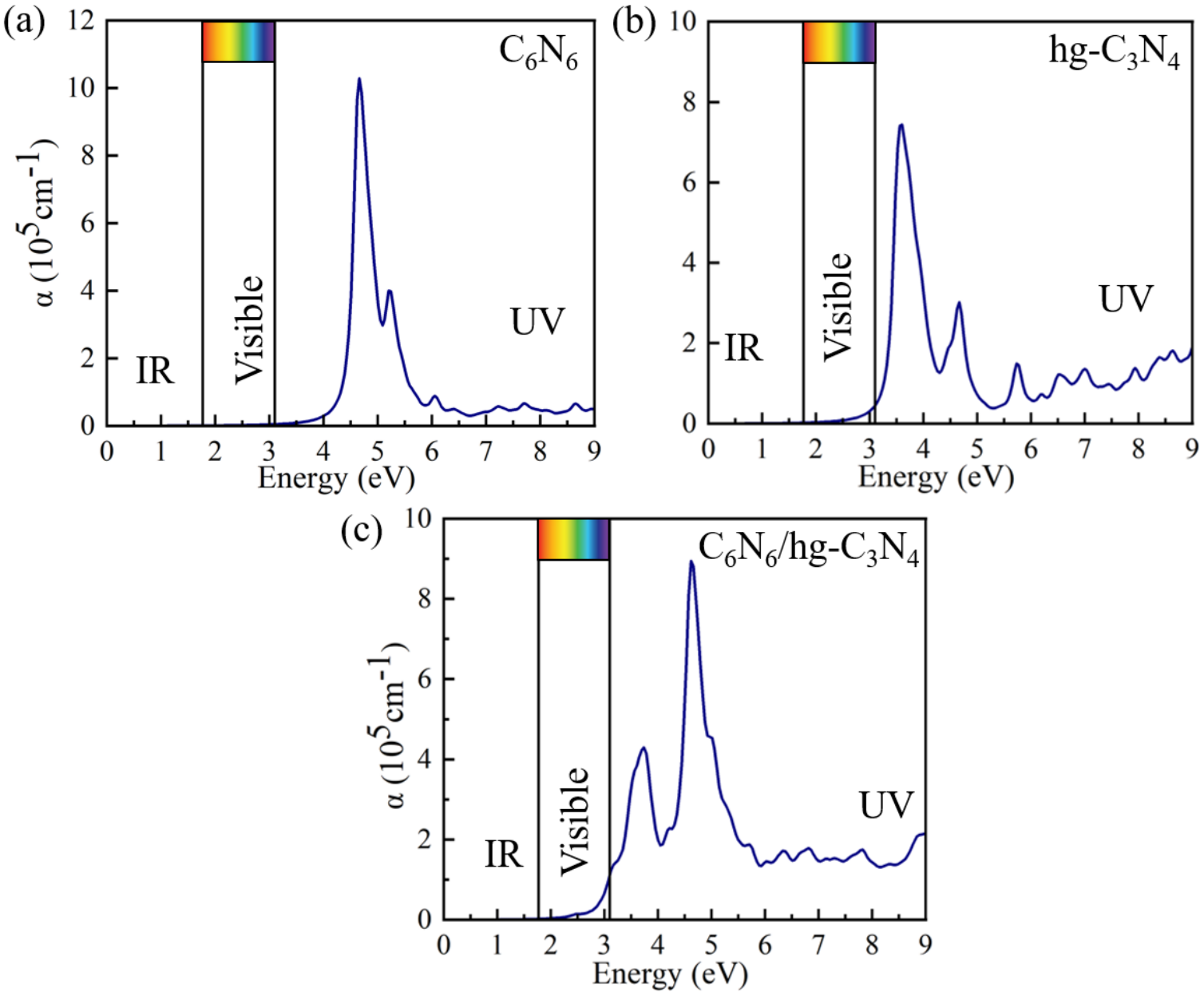}\caption{Calculated absorption coefficient ($\alpha$, cm$^{-1}$) for the
isolated monolayers (a) C$_{6}$N$_{6}$, (b) hg-C$_{3}$N$_{4}$,
and for the corresponding heterostructures (c) hg-C$_{3}$N$_{4}$/C$_{6}$N$_{6}$,
computed using the BSE on top of G$_{0}$W$_{0}$.}
\label{fig:BGW} 
\end{figure*}

Further, Fig. S8 of the SM~\cite{SI} shows the refractive index of the isolated
monolayers and the corresponding HTSs. For the GE monolayer (see Fig. S8(a) of the SM~\cite{SI}),
the static refractive index, $n(0)$, 1.83 is obtained. With increasing
photon energy, the refractive index rises in the IR region, reaching
a maximum value of 2.11 at 1.44 eV in the visible region. Beyond this
energy, $n(\omega)$ decreases within the visible region and subsequently
increases again in the UV region, exhibiting a peak at 3.22 eV. Further,
C$_{6}$N$_{6}$ monolayer (see Fig. S8(b) of the SM~\cite{SI}) exhibits a relatively
low $n(0)=1.27$, which increases gradually in the UV-Visible region,
reaching a peak of 1.57 at 4.38~eV, before decreasing for larger
energies. Moreover, hg-C$_{3}$N$_{4}$ monolayer (Fig. S8(c) of the SM~\cite{SI})
has $n(0)=1.40$, increasing towards the visible region and peaking
at 3.13 near 1.55 eV, and then decreases in the UV region. As far
as HTSs are concerned, a different refractive-index response is observed,
as shown in Figs. S7(d)-S7(f) of the SM~\cite{SI}. In the C$_{6}$N$_{6}$/GE
HTS (see Fig. S8(d) of the SM~\cite{SI}), the static refractive index exhibits
a very large value, $n(0)=21.04$, followed by a rapid decrease with
increasing photon energy. The refractive index drops sharply within
the IR region and attains values close to unity across the visible
and UV regions. A similar trend is noticed for the hg-C$_{3}$N$_{4}$/GE
HTS (see Fig. S8(e) of the SM~\cite{SI}), where an even larger value, $n(0)=27.64$,
is obtained. Upon increasing photon energy, $n(\omega)$ decreases
in the IR region and remains nearly constant with small fluctuations
throughout the visible and UV regions. In contrast, the C$_{6}$N$_{6}$/hg-C$_{3}$N$_{4}$
HTS (see Fig. S8(f) of the SM~\cite{SI}) exhibits a comparatively moderate refractive
index, with $n(0)=1.46$. The refractive index gradually increases
toward the visible region, reaching a maximum of 1.65 around 2.8-3.0~eV,
before decreasing in the UV region. Notably, this behavior closely
resembles that of the constituent monolayers.

Fig. S9 of the SM~\cite{SI} shows the calculated reflectivity spectra, $R(\omega)$,
of the isolated monolayers and their corresponding HTS, which exhibit
the trends complementary to those observed in the $L(\omega)$. The
GE monolayer (Fig. S9(a) of the SM~\cite{SI}) exhibits distinct reflectivity
peaks in the IR and UV regions. In contrast, the C$_{6}$N$_{6}$
(see Fig. S9(b) of the SM~\cite{SI}) and hg-C$_{3}$N$_{4}$ (see Fig. S9(c) of the SM~\cite{SI})
monolayers display comparatively weak reflectivity throughout the
spectrum, with broad maxima appearing in the UV range around 4-5~eV.
Upon HTS formation, a substantial enhancement of reflectivity is noticed
at low photon energies. In particular, the C$_{6}$N$_{6}$/GE (see Fig. S9(d) of the SM~\cite{SI})
and hg-C$_{3}$N$_{4}$/GE (see Fig. S9(e) of the SM~\cite{SI}) HTSs are characterized
by very high reflectivity in the IR region, approaching unity below
2~eV, followed by a rapid decrease as photon energy increases. Meanwhile,
the C$_{6}$N$_{6}$/hg-C$_{3}$N$_{4}$ HTS (see Fig. S9(f) of the SM~\cite{SI})
shows moderate reflectivity across the spectrum, characterized by
a distinct peak near 5.5~eV in the UV region. Overall, the relatively
low reflectivity values in the visible-UV range ($<$20-30\%) indicate
that these materials could be promising for optoelectronic and anti-reflective
coating applications.

\section{Exciton Binding Energy}
To understand the strength of the electron-hole interaction, the exciton binding energy, $E_{b}^{exc}$, is evaluated for the gapped systems, i.e, C$_{6}$N$_{6}$, hg-C$_{3}$N$_{4}$ monolayers, and the hg-C$_{3}$N$_{4}$/C$_{6}$N$_{6}$ HTS, using the G$_{0}$W$_{0}$+BSE approach, and the corresponding optical absorption spectra are presented in Fig. \ref{fig:BGW}. For the C$_{6}$N$_{6}$ and hg-C$_{3}$N$_{4}$ monolayers, the G$_{0}$W$_{0}$ quasiparticle band gaps are calculated to be 4.58 eV and 3.76 eV, respectively, whereas, for the hg-C$_{3}$N$_{4}$/C$_{6}$N$_{6}$ HTS, the value 3.55~eV is obtained. As far as the optical absorption onsets are concerned, within the RPA approach, they are 3.16 eV, 1.88 eV, and 1.89 eV for C$_{6}$N$_{6}$ (Fig.~\ref{fig:opti}(b)), hg-C$_{3}$N$_{4}$ (Fig.~\ref{fig:opti}(c)) monolayers, and their HTS, C$_{6}$N$_{6}$/hg-C$_{3}$N$_{4}$ (Fig.~\ref{fig:opti}(f)), respectively. Whereas the corresponding G$_{0}$W$_{0}$+BSE
optical onsets (see Fig.~\ref{fig:BGW}) are 3.57 eV, 2.62 eV, and 2.37 eV, clearly demonstrating the pronounced blue-shift in the absorption edge arising from the electron-hole interactions. Further, the exciton binding energies are computed as the difference between the G$_{0}$W$_{0}$ quasiparticle band gap and the optical gap obtained using the G$_{0}$W$_{0}$+BSE
approach for the respective system (see Eq.~\ref{excit}). The exciton binding energies thus obtained are 1.01 eV for the C$_{6}$N$_{6}$ monolayer, 1.14 eV for the hg-C$_{3}$N$_{4}$ monolayer, and 1.18 eV for the hg-C$_{3}$N$_{4}$/C$_{6}$N$_{6}$ HTS. Given the fact that in the HTS, the CBM and the VBM are localized on different monolayers (see Figs.~\ref{fig:EBS}(c) and (f)), the slightly higher $E_{b}^{exc}$ observed for the HTS reflects an enhanced electron-hole interaction
arising from the interlayer coupling and reduced dielectric screening. Further, these exciton binding energies are in good agreement with previously reported values for 2D materials, highlighting the pronounced excitonic character of the studied monolayers and heterostructure~\cite{yan2025janus}. The G$_{0}$W$_{0}$ quasiparticle band gap ($E_{g}^{\mathrm{G}_{0}\mathrm{W}_{0}}$), optical band gap ($E_{g}^{\mathrm{op}}$) obtained from BSE calculations on top of G$_{0}$W$_{0}$, and exciton binding energy ($E_{b}^{exc}$) for the C$_{6}$N$_{6}$, hg-C$_{3}$N$_{4}$ monolayers, and C$_{6}$N$_{6}$/hg-C$_{3}$N$_{4}$ HTS are summarized in Table~\ref{Table:GW}.

\begin{table}[ht]
\centering
\caption{Calculated $G_0W_0$ quasiparticle band gap ($E_g^{GW}$), optical band gap obtained from Bethe-Salpeter equation calculations on top of G$_{0}$W$_{0}$ ($E_g^{op}$), and exciton binding energy ($E_b^{exc}$) for C$_6$N$_6$, hg-C$_3$N$_4$ monolayers, and the C$_6$N$_6$/hg-C$_3$N$_4$ heterostructure.}
\label{Table:GW}
\begin{ruledtabular}
\begin{tabular}{lccc}
System & $E_g^{G_0W_0}$ (eV) & $E_g^{op}$ (eV) & $E_b^{exc}$ (eV) \\
\hline
C$_6$N$_6$ & 4.58 & 3.57 & 1.01 \\
hg-C$_3$N$_4$ & 3.76 & 2.62 & 1.14 \\
C$_6$N$_6$/hg-C$_3$N$_4$ & 3.55 & 2.37 & 1.18 \\
\end{tabular}
\end{ruledtabular}
\end{table}

\section{Conclusions}

In summary, first-principles density functional theory calculations
were employed to investigate the structural, electronic, mechanical,
and optical properties of C$_{6}$N$_{6}$, hg-C$_{3}$N$_{4}$, and
graphene monolayers, and their van der Waals heterostructures,
namely C$_{6}$N$_{6}$/GE, hg-C$_{3}$N$_{4}$/GE, and C$_{6}$N$_{6}$/hg-C$_{3}$N$_{4}$.
The study reveals that both semiconductor/metal heterostructures form
$n$-type Schottky contacts, which can be tuned to $p$-type Schottky or
Ohmic behavior by varying the perpendicular external electric field
and the interlayer coupling. In the C$_{6}$N$_{6}$/hg-C$_{3}$N$_{4}$
heterostructure, both the VBM and the CBM originate from different
constituent layers, giving rise to a type-II band alignment. Furthermore,
the band alignment can be tuned between type-I and type-II by an external
electric field and by changing the interlayer distance. Moreover,
the optical response of the considered monolayers and heterostructures
was investigated. The results on the monolayer were validated against
existing theoretical and experimental reports, while those on the
heterostructure represent novel theoretical predictions in the absence
of prior data. We also computed the exciton binding
energy for the gapped systems. The calculated exciton binding energies
of 1.01~eV, 1.14~eV, and 1.18~eV for the C$_{6}$N$_{6}$, hg-C$_{3}$N$_{4}$
monolayers, and their heterostructure (C$_{6}$N$_{6}$/hg-C$_{3}$N$_{4}$),
respectively, demonstrate pronounced electron-hole interactions.
Overall, our findings offer valuable insights into the tunability
of multifunctional properties in 2D van der Waals heterostructures,
paving the way for their potential applications in nanoelectronics,
optoelectronics, and other advanced technological fields. 

\begin{acknowledgments}
Poonam Sharma and Archana Sharma acknowledge the support from IIT Bombay in the form of an Institute Postdoctoral Fellowship. These calculations were performed on the \lq PARAM Rudra\rq~supercomputing facility at IITB, implemented by Centre for Development of Advanced Computing (C-DAC) and supported by the Ministry of Electronics and Information Technology (MeitY) and the Department of Science and Technology (DST), Government of India, under the National Supercomputing Mission (NSM).
\end{acknowledgments}

\bibliography{references}

\end{document}


\title{Supplemental Material for tuning electronic properties and Schottky contact in graphene-based van der Waals heterostructures by electric gating and interlayer coupling}
\author{Poonam Sharma}
\thanks{These authors contributed equally to this work.}

\author{Archana Sharma}
\thanks{These authors contributed equally to this work.}

\author{Alok Shukla}
\email{shukla@iitb.ac.in}
\affiliation{Department of Physics, Indian Institute of Technology Bombay, Mumbai
400076, India}

\maketitle

This supporting document contains the following entries: 
\subsection*{S1. Thermal Stability from \textit{ab initio} Molecular Dynamics (AIMD)}
\begin{figure*}[ht]
\includegraphics[width=1\linewidth]{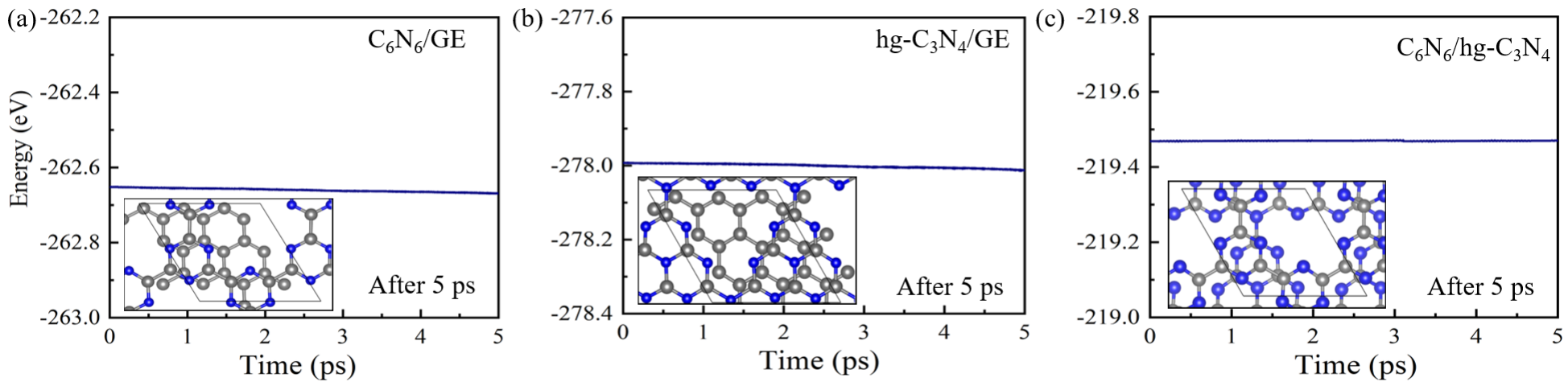}\caption{AIMD simulation for a time period of 5 ps at 300 K for (a) C$_{6}$N$_{6}$/GE,
(b) hg-C$_{3}$N$_{4}$/GE, and (c) hg-C$_{3}$N$_{4}$/C$_{6}$N$_{6}$
heterostructures, obtained using GGA.}
\label{fig:aimd} 
\end{figure*}

To examine the thermal stability of the considered heterostructures, \textit{ab initio} molecular dynamics (AIMD) simulations were performed for 5~ps at 300~K within the GGA framework. The time evolution of the total energy is shown in Fig.~\ref{fig:aimd} for the C$_6$N$_6$/GE, hg-C$_3$N$_4$/GE, and hg-C$_3$N$_4$/C$_6$N$_6$ heterostructures. In all cases, the absence of significant energy drift and structural distortion throughout the simulation confirms the thermal robustness and structural integrity of the proposed heterostructures at room temperature.

\subsection*{S2. Orientation-dependent in-plane elastic properties}
\begin{figure*}[ht]
\includegraphics[width=0.8\linewidth]{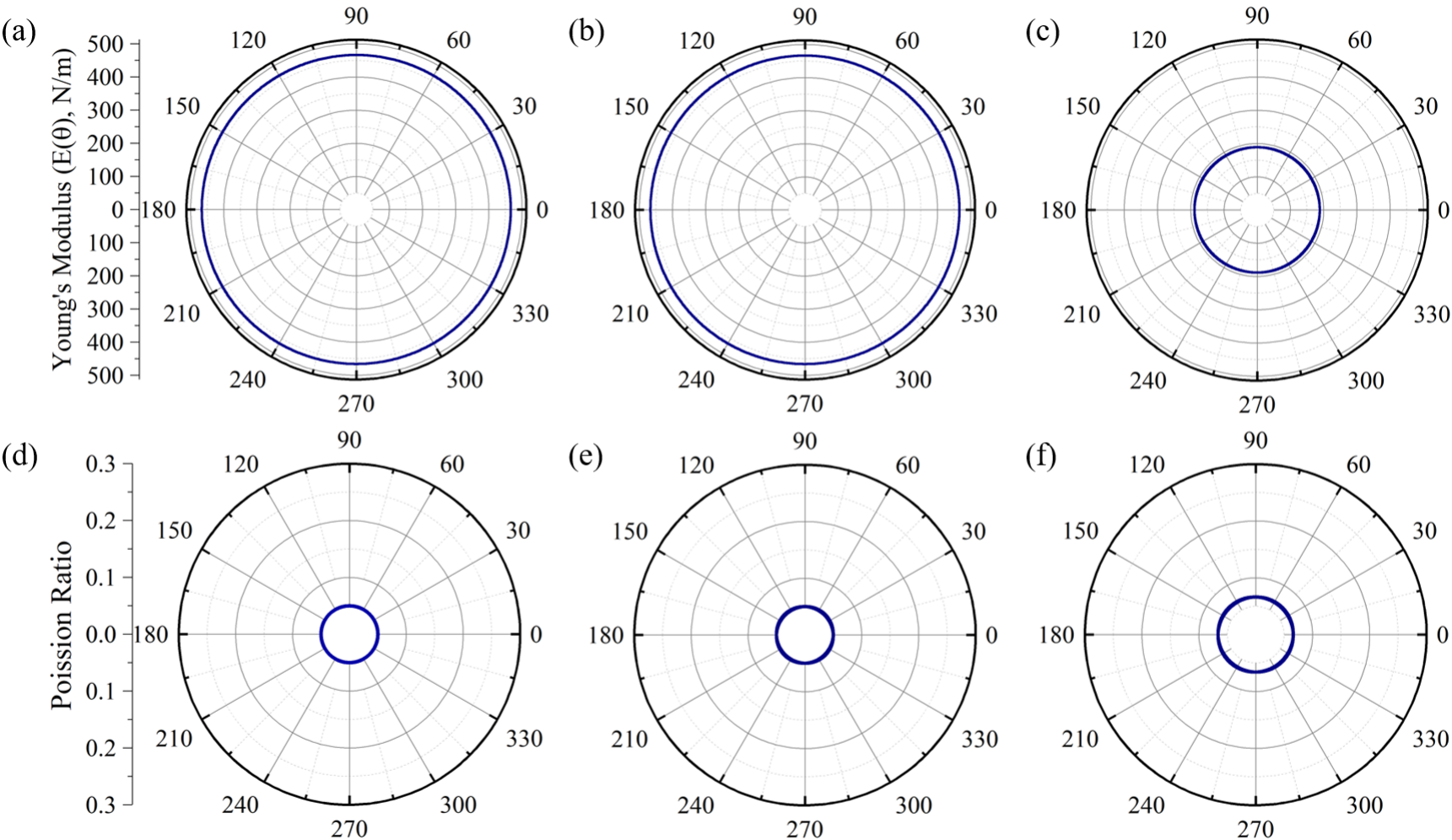}\caption{Calculated orientation-dependent (a)-(c) Young’s modulus ($E(\theta)$) and (d)-(f) Poisson’s ratio ($\nu(\theta)$) for the C$_6$N$_6$/GE, hg-C$_3$N$_4$/GE, and C$_6$N$_6$/hg-C$_3$N$_4$ heterostructures, respectively, using the GGA functional.}
\label{fig:s2} 
\end{figure*}

Fig.~\ref{fig:s2} shows the angular variation of Young’s modulus $E(\theta)$ and Poisson’s ratio $\nu(\theta)$ for the considered heterostructures. The nearly circular polar profiles of both quantities indicate an almost isotropic in-plane elastic response. The uniform distribution of stiffness with respect to the in-plane angle $\theta$, together with moderate Poisson’s ratio values, confirms the mechanical stability and strain tolerance of the heterostructures.

\newpage

\subsection*{S3. Electric-field dependent projected band structures}

\begin{figure*}[ht]
\includegraphics[width=0.8\linewidth]{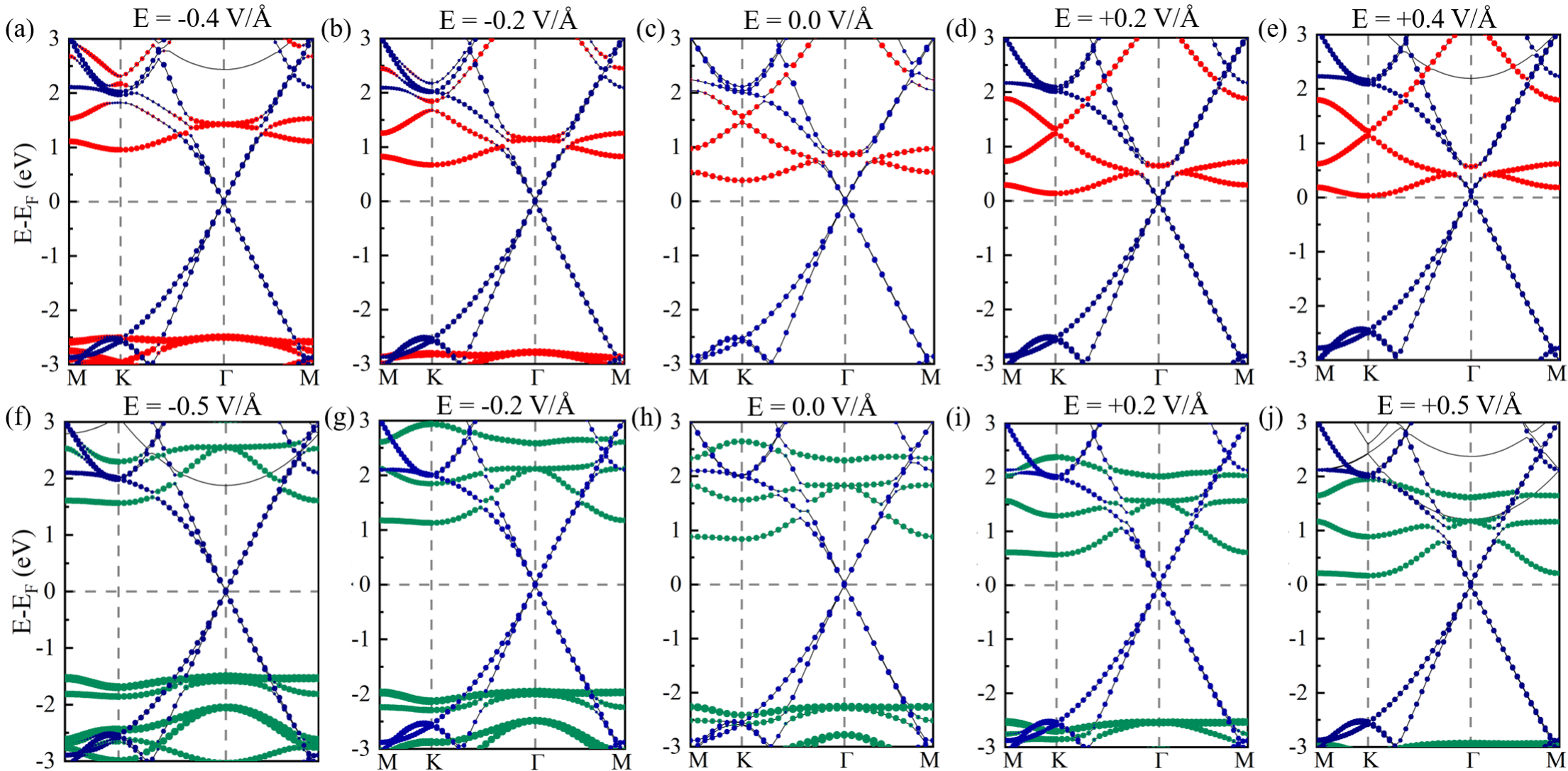}\caption{Projected band structures of the C$_{6}$N$_{6}$/GE heterostructure (top panels: (a)-(e)) and hg-C$_{3}$N$_{4}$/GE HTS (bottom panels:
(f)-(j)) under varying $E_{\perp}$, obtained using the HSE06 functional.
The changes in the positions of the conduction band minimum (CBM)
and valence band maximum (VBM) relative to the Fermi level (E$_{F}=0$)
highlight the influence of the applied $E_{\perp}$ on the electronic
properties of these HTSs. Blue color in all the band structure plots
indicates the contribution of the GE monolayer in the respective HTS,
whereas red and green represent the contributions from the C$_{6}$N$_{6}$
and hg-C$_{3}$N$_{4}$ monolayers, respectively.}
\label{fig:pro} 
\end{figure*}

Fig.~\ref{fig:pro} presents the projected band structures of the heterostructures under different perpendicular electric fields ($E_{\perp}$) calculated using the HSE06 functional. The gradual shift of the valence and conduction band edges with increasing $E_{\perp}$ reveals a clear field-induced modulation of the band alignment. The color-resolved projections show that the electronic states near the band edges progressively localize within different constituent layers, demonstrating the tunability of the interlayer charge distribution and electronic properties via an external electric field.

\subsection*{S4. Modulation of Schottky barrier height}
\begin{figure*}[ht]
\includegraphics[width=0.7\linewidth]{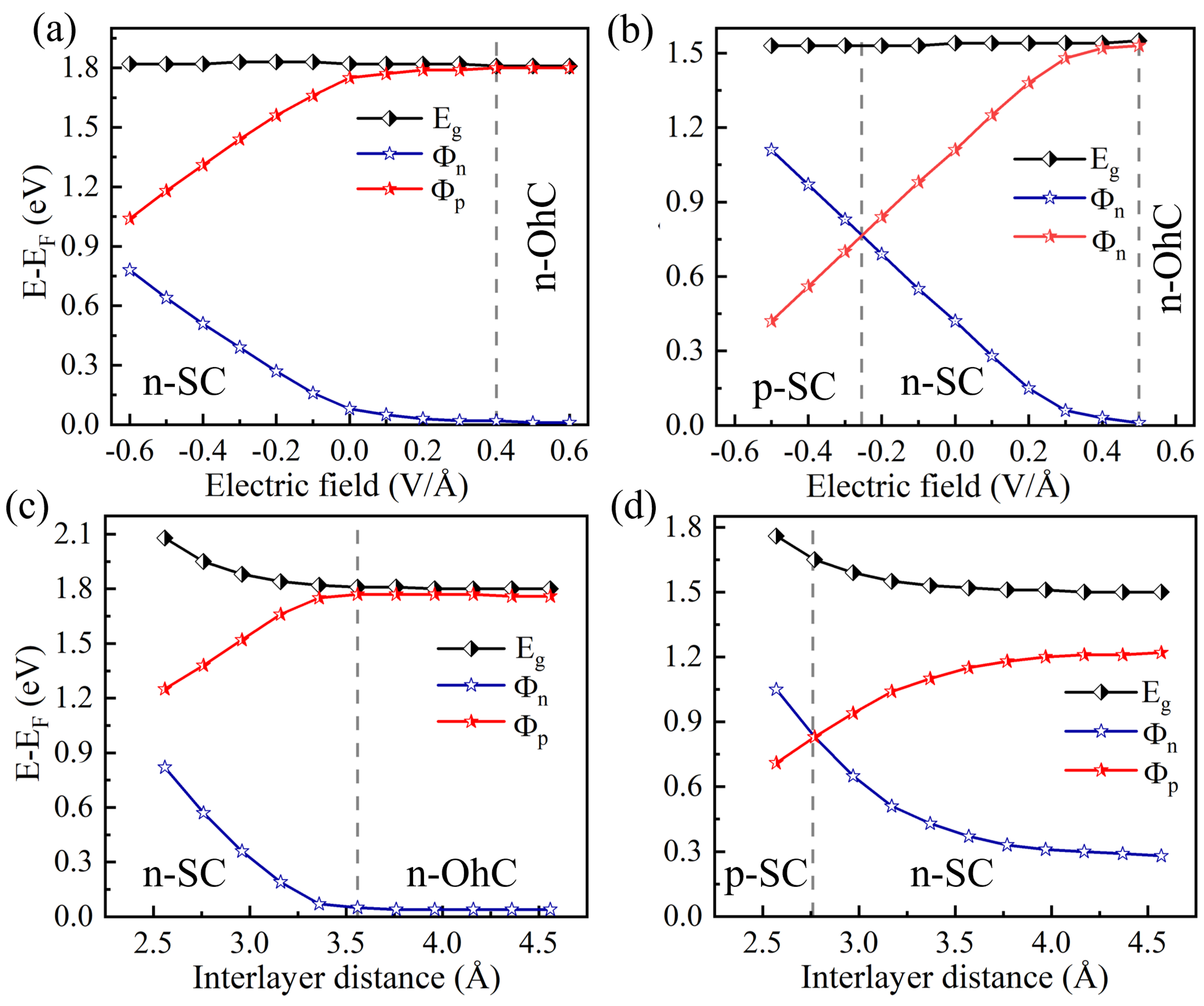}\caption{Variation of the Schottky barrier height (SBH) under an external perpendicular
electric field (E$_{\perp}$) for (a) C$_{6}$N$_{6}$/GE and (b)
hg-C$_{3}$N$_{4}$/GE heterostructures. (c), (d) Variation of the
SBH with interlayer distance for C$_{6}$N$_{6}$/GE and hg-C$_{3}$N$_{4}$/GE
heterostructures, respectively. All results are obtained using GGA.}
\label{fig:comb} 
\end{figure*}

Fig.~\ref{fig:comb} shows the variation of the Schottky barrier height (SBH) with applied perpendicular electric field ($E_{\perp}$) and interlayer distance for the C$_{6}$N$_{6}$/GE and hg-C$_{3}$N$_{4}$/GE heterostructures. A systematic change in SBH is observed with both external perturbations, indicating strong sensitivity of the interfacial electronic properties to electric field and layer separation. This behavior demonstrates the possibility of tuning the contact characteristics and carrier injection across the interface through electrostatic control and interlayer coupling.

\subsection*{S5. Interlayer-distance dependent projected band structures}
\begin{figure*}[ht]
\includegraphics[width=0.8\linewidth]{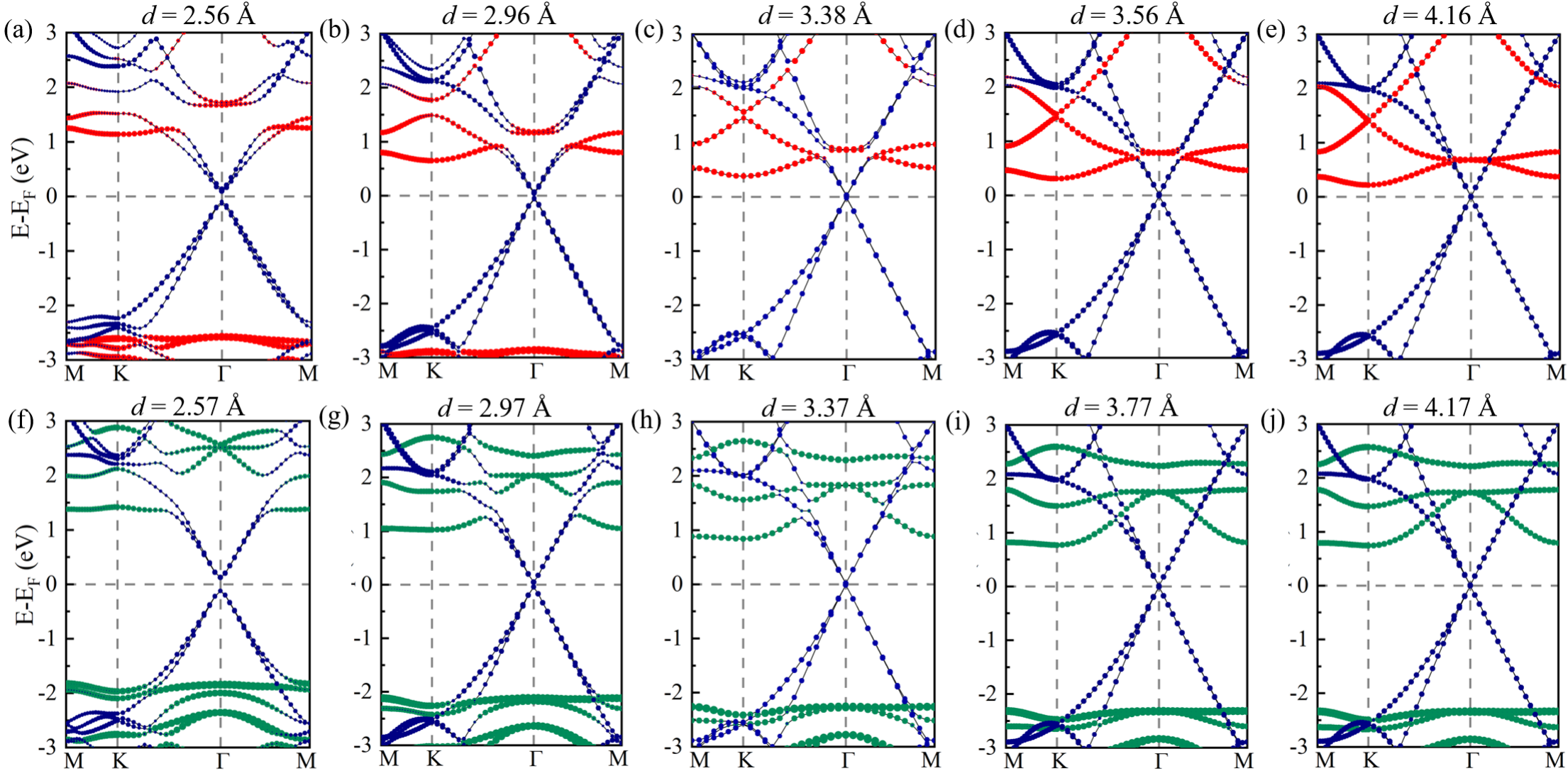}\caption{Projected band structures of the C$_{6}$N$_{6}$/GE heterostructure
(top panels: (a)-(e)) and hg-C$_{3}$N$_{4}$/GE HTS (bottom panels:
(f)-(j)) under varying interlayer distances (\emph{d}), obtained using the HSE06 functional. The changes in the positions of CBM and VBM relative to the Fermi level (E$_{F}=0$) highlight the influence of the interlayer distance on the electronic properties of these HTSs. Blue color in all the band structure plots indicates the contribution of the GE monolayer in the respective HTS, whereas red and green represent the contributions from the C$_{6}$N$_{6}$ and hg-C$_{3}$N$_{4}$
monolayers, respectively.}
\label{fig:verti} 
\end{figure*}

Fig.~\ref{fig:verti} presents the layer-projected band structures of the C$_{6}$N$_{6}$/GE and hg-C$_{3}$N$_{4}$/GE for different interlayer distances calculated using the HSE06 functional. A noticeable shift in the band edges is observed as the layer separation varies, indicating a strong dependence of the electronic structure on interlayer coupling. The color-resolved projections reveal the redistribution of band-edge states between the constituent layers, highlighting the sensitivity of band alignment to the interlayer distance.

\newpage
\subsection*{S6. Combined effect of electric field and interlayer distance}

\begin{figure*}[h]
\includegraphics[width=0.8\linewidth]{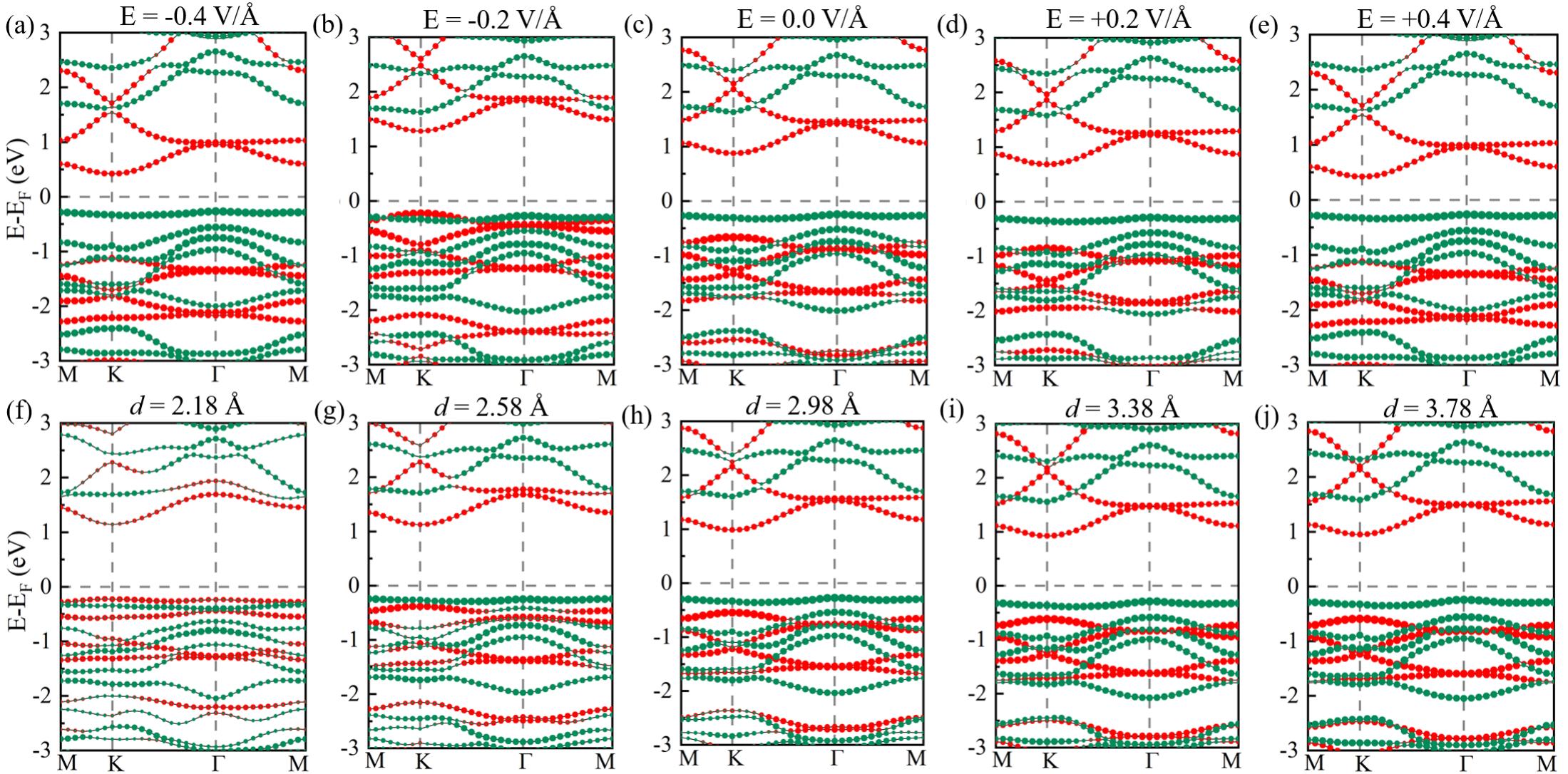}\caption{Projected band structures of the C$_{6}$N$_{6}$/hg-C$_{3}$N$_{4}$ heterostructure (top panels: (a)-(e)) under an external perpendicular electric field (E$_{\perp}$) and (bottom panels: (f)-(j)) under varying interlayer distances (\emph{d}), obtained using GGA. Red and Green
colors in all the band structure plots indicate the contribution from the C$_{6}$N$_{6}$ and hg-C$_{3}$N$_{4}$ monolayers, respectively.}
\label{fig:cont} 
\end{figure*}

Fig.~\ref{fig:cont} shows the projected band structures of the C$_{6}$N$_{6}$/hg-C$_{3}$N$_{4}$ heterostructure under varying perpendicular electric field ($E_{\perp}$) and interlayer distance. The evolution of the band edges under both perturbations indicates a tunable band alignment governed by interlayer coupling and electrostatic control. The color-resolved projections further reveal the redistribution of band-edge states between the two layers, confirming the sensitivity of the electronic structure to external and structural modifications.

\newpage
\subsection*{S7. Extinction coefficient and optical response}
\begin{figure*}[ht]
\includegraphics[width=0.7\linewidth]{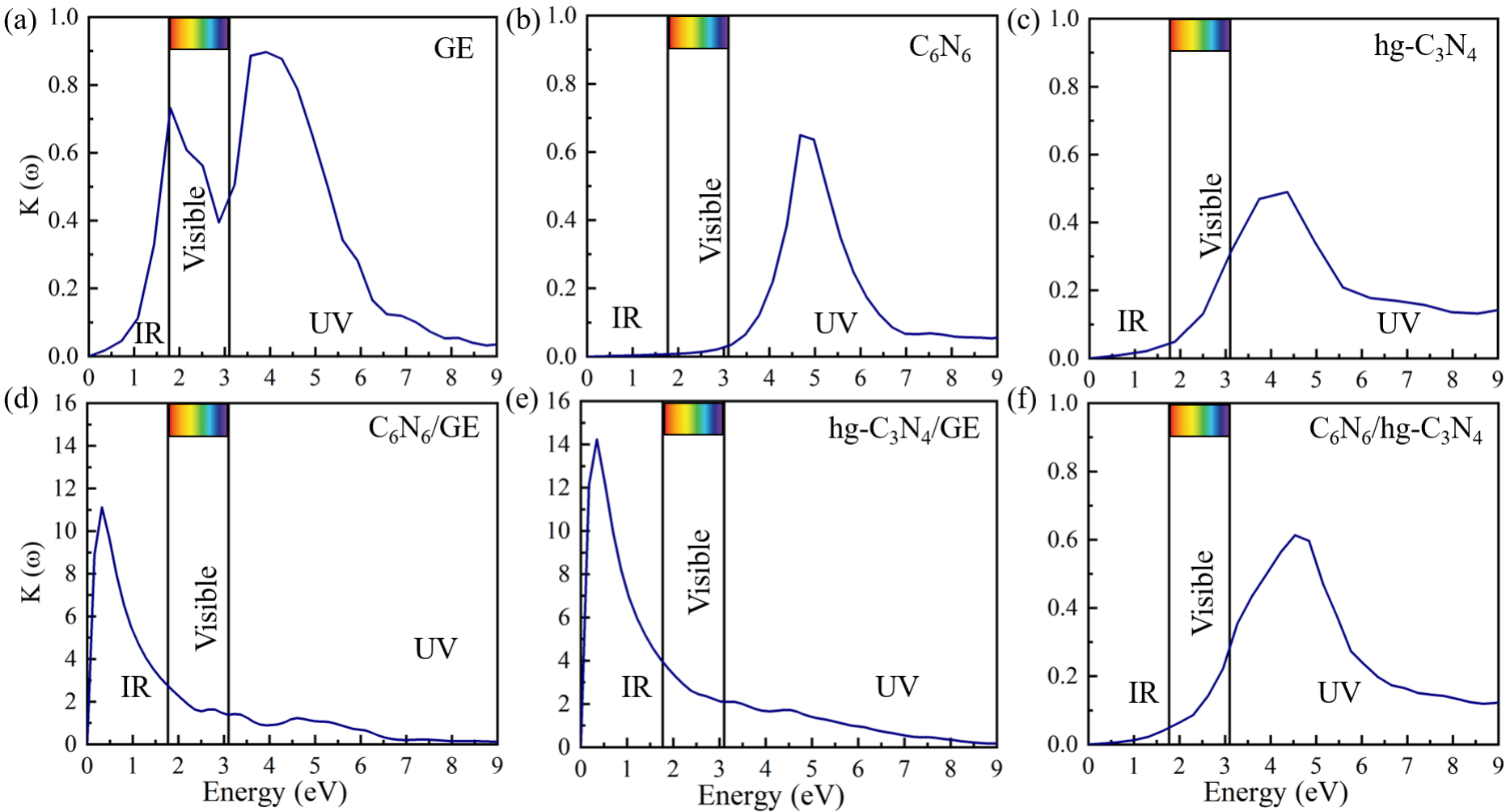}\caption{Calculated extinction coefficient ($k(\omega)$) for the isolated monolayers (a) GE, (b) C$_{6}$N$_{6}$, and (c) hg-C$_{3}$N$_{4}$, and for the heterostructures (HTSs) (d) C$_{6}$N$_{6}$/GE, (e) hg-C$_{3}$N$_{4}$/GE, and (f) hg-C$_{3}$N$_{4}$/C$_{6}$N$_{6}$, obtained within the RPA on top of GGA.}
\label{fig:S7} 
\end{figure*}

Fig.~\ref{fig:S7} shows the calculated extinction coefficient $k(\omega)$ for the individual monolayers and the corresponding heterostructures. A clear modification of the spectral features is observed upon forming the heterostructures, indicating the influence of interlayer interaction on the optical absorption characteristics and light-matter interaction in these systems.

\subsection*{S8. Refractive index and optical modulation}
\begin{figure*}[ht]
\includegraphics[width=0.7\linewidth]{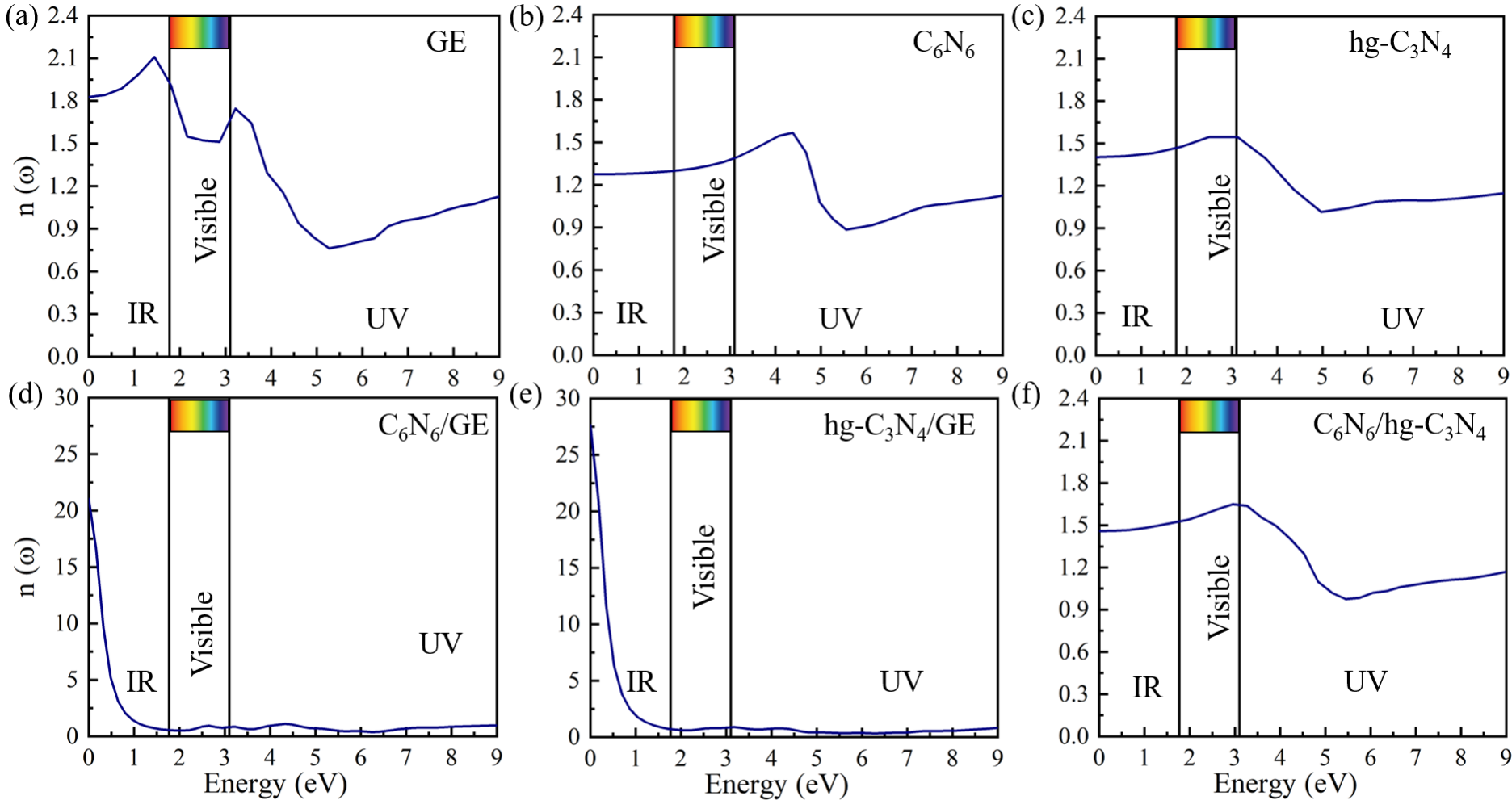}\caption{Calculated refractive index ($n(\omega)$) for the isolated monolayers (a) GE, (b) C$_{6}$N$_{6}$, and (c) hg-C$_{3}$N$_{4}$, and for the heterostructures (HTSs) (d) C$_{6}$N$_{6}$/GE, (e) hg-C$_{3}$N$_{4}$/GE, and (f) hg-C$_{3}$N$_{4}$/C$_{6}$N$_{6}$, obtained within the RPA on top of GGA.}
\label{fig:S8} 
\end{figure*}

Fig.~\ref{fig:S8} presents the calculated refractive index $n(\omega)$ for the individual monolayers and their corresponding heterostructures. Noticeable changes in the spectral profile upon forming the heterostructures indicate that interlayer interactions modify the optical dispersion and the overall light-propagation characteristics of the systems.

\subsection*{S9. Reflectivity}
\begin{figure*}[ht]
\includegraphics[width=0.7\linewidth]{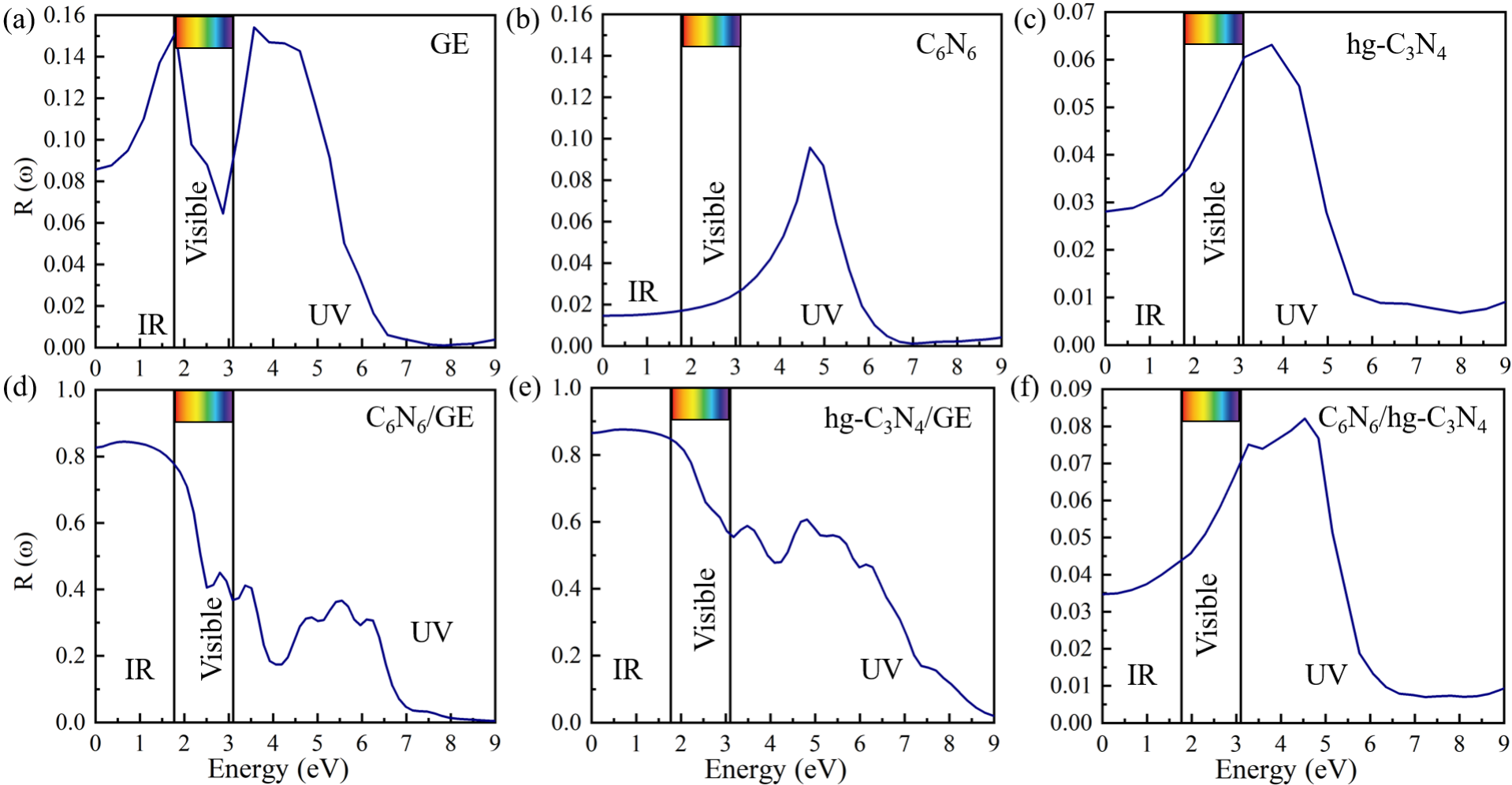}\caption{Calculated reflectivity ($R(\omega)$) for the isolated monolayers (a) GE, (b) C$_{6}$N$_{6}$, and (c) hg-C$_{3}$N$_{4}$, and for the heterostructures (HTSs) (d) C$_{6}$N$_{6}$/GE, (e) hg-C$_{3}$N$_{4}$/GE, and (f) hg-C$_{3}$N$_{4}$/C$_{6}$N$_{6}$, obtained within the RPA on top of GGA.}
\label{fig:S9} 
\end{figure*}

Fig.~\ref{fig:S9} shows the calculated reflectivity $R(\omega)$ for the isolated monolayers and the corresponding heterostructures. The variation in spectral features after forming the heterostructures highlights the influence of interlayer interactions on the reflection behavior and the overall optical response of the systems.

\bibliography{references}